\documentclass[preprint,
preprintnumbers,amsmath,amssymb,nofootinbib]{revtex4}

\usepackage{etex}
\usepackage{graphicx}
\usepackage{dcolumn}
\usepackage{bm}
\usepackage{amssymb,amsthm,amscd,amsbsy,array}\usepackage{amsmath,dsfont}
\usepackage{mathrsfs}
\usepackage[scr=rsfs,cal=boondox]{mathalfa}

\usepackage{graphics,xcolor}

\usepackage{color}
\newcommand{\red}{\textcolor{red}}  
\newcommand{\blue}{\textcolor{blue}}

\newcommand{\SU}{\mathrm{SU}}
\newcommand{\UU}{\mathrm{U}}
\newcommand{\half}{{\scriptstyle{\frac{1}{2}}}}
\def\2{{\half}}
\newcommand{\const}{\mathop{\rm const}\nolimits}
\def\parag{\hfil\break} 
\def\kikezd{\parag\underbar}

\def\su{{\mathrm{su}}}
\def\p{{\partial}}

\def\bsigma{{\bm{\sigma}}}

\def\hx{{\hat{\bm{r}}}}

\def\hPsi{{\widehat{\Psi}}}
\def\bx{{\bm{x}}}
\def\vr{{\bm{r}}}

\def\beq{\begin{equation}}
\def\eeq{\end{equation}}
\def\beqa{\begin{eqnarray}}
\def\eeqa{\end{eqnarray}}

\def\barray{\left(\begin{array}}
\def\earray{\end{array}\right)}
\def\barraynb{\begin{array}}
\def\earraynb{\end{array}}
\def\IR{{\mathds{R}}} 
\def\IZ{{\mathds{Z}}}

\def\IS{{\mathds{S}}} 

\def\SO{{\rm SO}}
\def\smallover#1/#2{\hbox{$\textstyle\frac{#1}{#2}$}} %
\def\vx{{\vec{x}}}
\def\vQ{{\vec{Q}}}

\def\vJ{{\vec{J}}}
\def\vL{{\vec{L}}}

\def\vPhi{{\vec{\Phi}}}

\def\vp{{\vec{p}}}
\def\vpi{{\vec{\pi}}}
\def\vn{{\hat{z}}}

\def\vA{{\vec{A}}}

\newcommand{\tr}{\mathrm{tr}}
\newcommand{\diag}{\mathrm{diag}}

\newcommand{\fG}{\mathfrak{G}}
\newcommand{\fK}{\mathfrak{K}}
\newcommand{\fk}{\mathfrak{k}}

\newcommand{\fF}{\mathfrak{F}}
\newcommand{\fa}{\mathfrak{a}} 

\newcommand{\II}{\mathds{I}}

\newcommand{\cF}{\mathcal{F}}

\newcommand{\homega}{{\widehat{\Psi}}}

\newcommand{\cO}{{\mathcal{O}}}

\def\smallcirc{{\raise 0.5pt \hbox{$\scriptstyle\circ$}}}
\def\aand{{\quad\text{\small and}\quad}}
\def\for{{\quad\text{\small for}\quad}}
\def\when{{\quad\text{\small when}\quad}}
\def\st{{\quad\text{\small such \ that}\quad}}
\def\where{{\quad\text{\small where}\quad}}

\def\oor{{\quad\text{\small or}\quad}}
\def\ie{{\;\text{\small i.e.}\;}}
\def\ie,{{\;\text{\small i.e.,}\;}}

\def\benu{\begin{enumerate}}
\def\eenu{\end{enumerate}}
\def\bitem{\begin{itemize}}
\def\eitem{\end{itemize}}

\def\besub{\begin{subequations}}
\def\esub{\end{subequations}}

\def\?{{\,\gb{\fbox{\texttt{??}}\;}}\,}

\def\Rarrow{{\quad\Rightarrow\quad}}

\usepackage[colorlinks=true, pdfstartview=FitV, linkcolor=blue, citecolor=blue, urlcolor=blue]{hyperref} 
\newcommand{\cyan}{\textcolor{cyan}}
\newcommand{\magenta}{\textcolor{magenta}}

\newcommand{\purple}{\textcolor[rgb]{0.5,0.0,0.5}}
\newcommand{\gb}{\quad\colorbox{green}}

\newcommand{\dgreen}{\textcolor[rgb]{0,0.5,0}}

\newenvironment{redtext}{\color{red}}
{\ignorespacesafterend}
\newenvironment{bluetext}{\color{blue}}{\ignorespacesafterend}
\newenvironment{greentext}{\color{green}}{\ignorespacesafterend}
\newenvironment{magentatext}{\color{magenta}}{\ignorespacesafterend}
\newenvironment{cyantext}{\color{cyan}}{\ignorespacesafterend}
\newenvironment{orangetext}{\color{orange}}
{\ignorespacesafterend}

\newcommand{\bmagenta}{\begin{magentatext}}
\newcommand{\emagenta}{\end{magentatext}}
\newcommand{\bcyan}{\begin{cyantext}}
\newcommand{\ecyan}{\end{cyantext}}
\newcommand{\bblue}{\begin{bluetext}}
\newcommand{\eblue}{\end{bluetext}}
\newcommand{\bred}{\begin{redtext}}
\newcommand{\ered}{\end{redtext}}

\newcommand{\bgreen}{\begin{greentext}}
\newcommand{\egreen}{\end{greentext}}
\newcommand{\borange}{\begin{orangetext}}
\newcommand{\eorange}{\end{orangetext}}

\numberwithin{equation}{section}

\let\ssection=\section
\renewcommand{\section}{\setcounter{equation}{0}\ssection}

\newcommand{\bigbox}[1]{\fbox{%
\rule[-20pt]{0pt}{45pt}$\;\;\displaystyle{#1}\;\;$}
}
\newcommand{\medbox}[1]{\fbox{%
\rule[-10pt]{0pt}{25pt}$\;\;\displaystyle{#1}\;\;$}%
}

\newcommand{\NABA}{Non-Abelian Aharonov-Bohm Effect\;}

\newcommand{\AB}{Aharonov-Bohm\;}

\newcommand{\NA}{non-Abelian \;}

\newcommand{\NAPF}{Non-Abelian Phase Factor\;} 

\newcommand{\NIPF}{non-integrable phase factor\;}

\usepackage{amsthm}
\newtheorem{thm}{Theorem}[section]
{Corollary}[section]
\newtheorem{definition}{Definition}[section]

\begin{document}

\preprint{{arXiv:2402.13883v2 [hep-th]}
}

\title{Isospin precession \\ in  \\ non-Abelian Aharonov-Bohm scattering \footnote{Dedicated to the memory of Professor Tai-Tsun Wu (1933-2024).}
}

\author{
P.-M. Zhang$^{1}$\footnote{mailto:zhangpm5@mail.sysu.edu.cn},
and
P. A. Horv\'athy$^{2}$\footnote{mailto:horvathy@lmpt.univ-tours.fr}
}

\affiliation{
$^1$ School of Physics and Astronomy, Sun Yat-sen University, Zhuhai 519082, (China)
\\
$^2$ Institut Denis-Poisson CNRS/UMR 7013 - Universit\'e de Tours - Universit\'e d'Orl\'eans Parc de Grammont, 37200, Tours, (France)\\
}
\date{\today}

\begin{abstract} 
The concept of pseudoclassical isospin 
 is illustrated by the non-Abelian Aharonov-Bohm effect proposed by Wu and Yang in 1975. The spatial motion is free however the isospin precesses  when the enclosed magnetic flux and the incoming particle's isosopin are not parallel. 
The non-Abelian phase factor $\mathfrak{F}$ of Wu and Yang acts on the isospin as an S-matrix. The scattering becomes side-independent when the enclosed flux is quantized, ${\Phi}_N=N\Phi_0$ with $N$ an integer. The gauge group $SU(2)$ is an internal symmetry and generates  conserved  charges only when the  flux is quantized, which then splits into two series: for $N=2k$ $SU(2)$ acts trivially but for $N=1+2k$ the implementation is twisted. The orbital and the internal angular momenta are separately conserved. The double rotational symmetry is broken to  $SO(2)\times SO(2)$ when $N$ odd. For unquantized flux there are no internal symmetries, the charge is not conserved and protons can be turned into neutrons.
\end{abstract}

\maketitle

\tableofcontents

\newpage

\section{Introduction: the vector potential in gauge theory}\label{Intro}
The vector potential has long been thought of as an auxiliary quantity until Aharonov and Bohm  pointed out that it can produce an observable effect \cite{AharonovBohm}, shown schematically on FIG. \ref{ABsetup}~:
\begin{figure}[h]
\includegraphics[scale=.7]{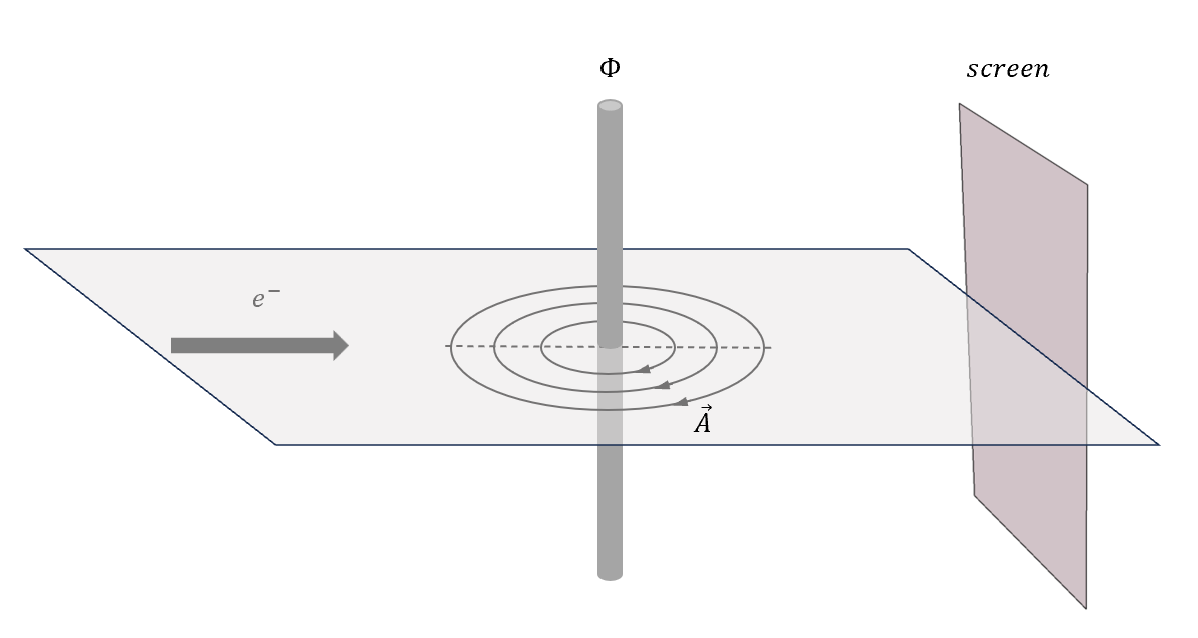}\vskip-4mm
\caption{\textit{\small Schema of the Abelian \AB 
experiment \cite{WuYang75}.} 
\label{ABsetup}
}
\end{figure}
electrically charged 
\emph{classical} particles move freely outside a cylinder
carrying a magnetic flux, however the curl-free vector potential $A=A_{\mu}dx^{\mu}$ allows for a \emph{quantum effect}. The  experimental confirmation of the \AB prediction \cite{Chambers, Mollen,Osakabe,Peshkin89,Tonomura87,Tonomura89,TonomuraToday} led  Wu and Yang to argue in their seminal paper \cite{WuYang75} that~:

\begin{quote} \textit{\narrower 
 (a) The field strength $F_{\mu\nu}$ under-describes (b) the phase $\oint\!{A_\mu} dx^{\mu}$ over-describes electromagnetism, 
 What provides a complete description that is neither too much nor too little is,
 \beq
 \fF^{AB}=
\exp\left[\,-i\!\oint A_{\mu}dx^{\mu}\right]\,. 
\label{ABPF}
\eeq
[\dots]
Electromagnetism is the gauge-invariant manifestation of a nonintegrable phase factor.}
\end{quote} 

Here we took unit electric charge; $\fF^{AB}= e^{-i\,\Phi}$ where
 $\Phi=\oint\! A_{\mu}dx^{\mu}$ is the magnetic flux enclosed  by a loop which goes  around the cylinder once counter-clockwise \footnote{In geometric language, such a loop generates the first homotopy group $\pi_1\simeq \IZ$ of the space from which the cylinder has been removed \cite{Schulman71,Schulman81,Laidlaw,Dowker72,Dowker79,Horvathy88,Horvathy79,Horvathy80}. All values of the  $\fF$ span the holonomy group \cite{Kobayashi,Husemoller}.}. 
Dropping the $z$ direction we  choose the gauge where 
$
A = A_\mu dx^{\mu}  = {\alpha}\, d{\vartheta}\,. 
$
Shifting ${\alpha}$ by an integer, ${\alpha} \to {\alpha}+N$, shifts the  magnetic flux,   
 $\Phi \to \Phi + {2N}\pi$, but yields identical Wu-Yang factors,
\beq
\Phi^{AB} = -2\pi{\alpha}-2N\pi \Rarrow {\fF}_{\alpha}^{AB}=
e^{i2{\pi}{\alpha}}\,. 
\label{ABflux}
\eeq
\noindent 
Two vector potentials are therefore physically equivalent whenever their fluxes differ by an  \emph{even multiple} of $\pi$ \cite{AharonovBohm,Chambers, Mollen}, 
$ 
\Phi^{(1)}-\Phi^{(2)} = {2N}\, \pi, \; N=0,\, \pm 1,\dots
$ 
\cite{Peshkin89,Tonomura87,Tonomura89,TonomuraToday}.
\goodbreak

The non-Abelian generalisation of electromagnetism  was proposed by Yang and Mills (YM) in 1954  \cite{YangMills}. 
Twenty years later, Wu and Yang \cite{WuYang75} 
underlined the r\^ole of the non-Abelian generalisation of the phase factor \eqref{ABPF},  
\beq
{\fF}^{NA} 
 =\exp[-\Phi] =\exp\big[-\oint\! A_{\mu}dx^{\mu}\big] \,.
\label{WYLoop}
\eeq
This expression, which will hence be referred to as the \emph{Wu-Yang factor}, looks as the Abelian expression \eqref{ABflux} except for that $A_{\mu} =\big(A_{\mu}^a\big)$ is the Lie algebra valued Yang-Mills potential,  and path-ordering is understood \footnote{Greek indices $\mu,\,\nu$ etc refer to space-time  with $i,\,j$ etc space indices. Latin indices $a,\, b$ etc refer to the gauge group.}. 

To verify their proposal Wu and Yang suggested, by analogy with the electromagnetic case, an \emph{experimentum crucis}. 
 Quoting again from  \cite{WuYang75}, sec. VII, p. 3855~:
 
\begin{quote}\textit{\narrower 
 \dots many theorists \dots believe that $\SU(2)$, gauge fields do exist. However, so far there is no experimental proof, since conservation of isotopic spin only suggests, and does not require, the existence of an isotopic spin gauge field. 
 What kind of experiment would be a definitive test of the existence of an isotopic spin gauge field? A generalized \underline{Bohm-Aharonov experiment} would be.
  One constructs the cylinder of material for which the total $I_z$ spin is not zero, e.g., a cylinder made of heavy elements with a neutron excess. One spins the cylinder around its axis, 
  \dots setting  up a \underline{``magnetic`` flux} inside the cylinder \underline{along the $I_z$ ``direction''}} (see Fig.\ref{ABsetup}). 
\end{quote}

\begin{quote} 
 \textit{\narrower \dots 
 one imagines a rotating a cylinder which has 
 an  average $\langle\vec{I}\rangle$ 
 which is [\,\dots\,] 
  not in the $I_z$ direction [with] a  magnetic flux would then be set up 
which is in a ``direction`` other than $I_z$. \underline{Scattering a beam of protons would then} \underline{ produce some neutrons as well as protons}. 
This implies, of course, that there is electric charge transfer between the beam and the cylinder together with the gauge field around it.
 }
\end{quote}
\goodbreak
 
In this paper we argue  that while there is no classical  electromagnetic \AB effect, however 
in the \NA case a ``shadow'' of the quantum effect, namely  \emph{isospin precession} \cite{WuYang75,NABA82,MSW,Jackiw86,KernerFest} can be recovered
at the \emph{(pseudo-) classical level}.  
 The difference with the Abelian case comes from that a  particle with \NA  structure carries an additional, internal degree of freedom represented by an \emph{isospin vector} $\vec{Q}=(Q^a)$ [or equivalently, by a matrix  in the Lie algebra, $\su(2)$, of the gauge group] which couples the particle to the YM field. The clue is that, unlike the electric charge, the isospin has \emph{non-trivial dynamics} rather than being a mere constant \cite{Kerner68,Wong70,Trautman70, Cho75, Witten81,Balachandran76, Balachandran77, Sternberg77, Sternberg78, DuvalAix79,DH82}.
 Then our first result says: 

\begin{thm}
The isospin of a (pseudo) classical particle with internal stucture in the \NA \AB field precesses around the non-Abelian flux in isospace.
\label{isospinperec}
\end{thm}

\smallskip
The original Wu-Yang proposal does not appear practical \footnote{Zeilinger et al.  \cite{Zeilinger} did actually scatter a neutron beam on a rotating rod of ${}^{138}$U, setting up an upper limit  $10^{-15}$ on the strength of the \NA effect when compared to its electromagnetic counterpart.}. 
 Therefore the Wu-Yang proposal has long be view as a Gedankenexperiment and the study was focused at particles coupled to a pure $\SU(2)$ Yang-Mills field with no electromagnetic interaction~\footnote{Ignoring the electric charge is reminiscent of von Neumann's ``dry water'', though.}.  
However, as emphasised by Yang and Mills already \cite{YangMills}, 
 
\begin{quote}\textit{\narrower
``\dots when electro-magnetic interactions can be neglected, 
[\dots] 
the orientation of the isotopic spin is of no physical significance.  The \underline{differentiation between a neutron and a proton} is then a 
  purely arbitrary process.  
However, %
[\dots] this arbitrariness is subject to the following limitation: 
[\dots] \underline{once one chooses what to call a proton, what a neutron at one space-time point,} one is then \underline{not free} to make any choices at other space-time points.''
}
\end{quote}

However the Wu-Yang suggestion that incoming protons could be turned into neutrons raises deep questions, as it leads  to conflict with electric charge conservation \cite{WuYang75,NABA82,HR-PRD,HR-color,NABAPRD}~\footnote{{*}The question of the referee of \cite{NABA82} 
 about  charge conservation which turned out to be related to the so-called ``color problem'' 
\cite{Schwarz82,HR-PRD,NelsonMano83,MarmoBala82,NelsonColeman84,HR-color} led subsequently to \cite{NABAPRD}. }.
When should we call a particle a proton or a neutron ? 
Following Yang and Mills, we posit~:
\begin{definition} 
The definition of protons and neutrons should be \emph{gauge covariant}. A \emph{proton} must have \emph{charge $1$} and a \emph{neutron} must have \emph{charge $0$}.
\label{protonneutrondef}
\end{definition} 

Putting the Yang-Mills argument above into practice, we attach a \emph{reference frame} to each point of spacetime such that satisfies the critera above and then allows us to distinguish protons and neutrons.

Our paper is organized as follows. 
After outlining  the general theory, we focus our attention at the pseudoclassical description of the \NABA\!. 

Then in sec.\ref{NABAIntSymSec}  we derive 
 conserved charges, associated with internal symmetries. 
Remarkably, \NA \AB field admits
two types of the latter: a trivial one and a ``twisted'' one, which arise when the enclosed \emph{flux
is quantized in even resp. in odd integers}, see eqn. 
\eqref{NAfluxquant} \cite{HR-PRD,HR-color,NABAPRD}.

Important conceptual and experimental progress has been made since those times, though
 \cite{WILCZEKCOLEMAN,Wilczek89, Alford90, Preskill90, Brandenberger93, MSW,Jackiw86,KernerFest,Bright15,Hosseini16,Chatterjee15,Cserti}. 
The developments \cite{Fruchart,Osterloh,Goldman,Jacob,Dalibard,YChen,YYang,YBiao,YangYang,NABA-PR} which concern \emph{effective non-Abelian gauge fields} change the assessment dramatically, opening the way to laboratory testing.
%
\goodbreak

\section{The Wu-Yang phase factor}\label{YMSec} 

Hence we limit our investigations to $G=\SU(2)$ Yang-Mills fields.  
The Lie algebra $\fG =\su(2)$ is generated by the  Pauli matrices,
\beq
\sigma_1 = {\tiny \barray{cc}0 &1 \\ 1 &0\earray}\,,
\;\;
\sigma_2 = {\tiny \barray{cc}0 &-i \\ i &0\earray}\,,
\;\;
\sigma_3 = {\tiny \barray{cc}1 &0 \\ 0 &-1\earray}\,,
\label{sigmas}
\eeq 
whose commutation relations are
$ 
[\sigma_a,\sigma_b]= 2i\epsilon_{abc}\sigma_c\,.
$ 
The correspondance between 3-vectors $\vQ$ and $\su(2)$ matrices $Q$ is,
\beq
\vec{Q} \equiv {\tiny\barray{c}Q^1 \\ Q^2\\ Q^3\earray}
\;\;\Leftrightarrow \;\;
Q^1\,\frac{\sigma_1}{2i}+Q^2\,\frac{\sigma_2}{2i}+Q^3\,\frac{\sigma_3}{2i}
 =
\smallover{1}/{2i}{\barray{cc}Q^3 &Q^1-iQ^2 \\[6pt]
 Q^1+iQ^2 &-Q^3\earray}\equiv Q\,.
 \label{Qvecmatrix}
\eeq
The vector and matrix forms will be used alternatively.
The gauge group $G$ is implemented according to,
\beq
g \,\cdot \vec{Q} \quad\ie,\quad  (g^{1/2})^{-1} Q\, g^{1/2}\,,  
\label{vecmatimp}
\eeq
where $g^{1/2}$ is square-root of the $\SU(2)$ matrix $g=(g_{ab})$ and $g\cdot \vQ$ means the action of $\SO(3) = \SU(2)/\IZ_2$ matrix on the 3-vector $\vQ$. The Yang-Mills vector potential and field strength \cite{YangMills} are given by,
$
A \equiv A_\mu dx^{\mu} 
$ and
$
F=\half F_{\mu\nu} dx^{\mu}\wedge dx^{\nu}\,,
$
respectively, where $A_\mu=(A_{\mu}^a)$ and 
$F_{\mu\nu}=(F_{\mu\nu}^{a})$ are $\fG =\su(2)$ matrices.  
$
A_\mu \to g^{-1} A_\mu \, g + g^{-1} \partial_\mu g\,,
\;
F_{\mu\nu} \to g^{-1}F_{\mu\nu} \, g\,
$
under $\SU(2)$-valued gauge transformations \cite{YangMills}.

\smallskip
We focus henceforth our attention
at the \NABA\!. A \NA  flux is confined to a thin cylinder  outside of which $F_{\mu\nu}=0$. Dropping the irrelevant $z$ coordinate, we consider a flat $\SU(2)$ Yang-Mills potential $A=A_{\mu}dx^{\mu}$, $DA=0$, on the punctured plane $\IR^2\setminus \{0\}$\,.  
Such Yang-Mills fields are characterized by the non-Abelian generalization of the non-integrable phase factor \eqref{ABPF}, defined, for an arbitrary path $\gamma$, as
\beq
\cF^{NA}(\gamma) =
 \exp\big[-\int_\gamma \!A_{\mu}dx^{\mu}\big] \in \SU(2)\,, 
\label{NIPFeq}
\eeq
with path-ordering understood \cite{WuYang75}. 
 Was the gauge field strength non-zero, $F_{\mu\nu}\neq0$,  the result would depend on the path. 
However in the \NA \AB case the non-Abelian version of Stokes' theorem \cite{GoddardOlive} implies that, owing to $DA=0$, \eqref{NIPFeq} is a homotopy invariant \cite{NABAPRD}, \ie, it takes the same values when $\gamma$ is continuously deformed while keeping its end points fixed.
For a generating loop  of $\pi_1 \approx \IZ$ of the punctured plane which  starts and ends at some point $x_0$ \eqref{NIPFeq}  
depends on the choice of $x_0$~:
choosing a different $x_0$ results by conjugation by a fixed element of the gauge group. Keeping this fact in mind, we shall write simply ${\fF}=\exp[-\Phi] =\exp[-\oint\! A_{\mu}dx^{\mu}]$ as in \eqref{WYLoop}. The suffix ``NA'' will be dropped henceforth and the non-Abelian character tacitly assumed. 

In the gauge we shall call diagonal the non-Abelian vector potential is in polar coordinates $(r,\vartheta)$,
\beq 
A_r = 0,\quad 
A_{\vartheta}= {\fa}\,\frac{\sigma_3}{2i}\,.
\label{diagpot}
\eeq 
The generated non-Abelian flux\footnote{Our non-Abelian flux $\Phi$ is Wu-Yang's $I$ \cite{WuYang75}.} and Wu-Yang factor are, 
\besub
\begin{align}
\Phi
 = & \;\oint\!A_i dx^i 
= -i\pi{\fa}\,\sigma_3\,,
\label{NAflux}
\\[4pt]
{\fF} = & \;\exp[-\Phi] = \;
\barray{lr}e^{{i}\pi{\fa}} & 0\\ 0 &e^{-{i}\pi{\fa}}
\earray\,,
\label{NAWYfactor}
\end{align}
\label{flux+factor} 
\esub
respectively.
 It follows that gauge potentials which differ by an \emph{even} integer,
\beq
{\fa} \aand  {\fa} + 2k,
\quad k = 0, \pm 1,\dots
\label{a2k}
\eeq
\ie, whose \NA fluxes are related as
\beq
\Phi \aand 
\Phi - 2k\pi{i}\,\sigma_3\,,
\label{fluxdiff}
\eeq
respectively, define identical Wu-Yang factors and are therefore physically equivalent.
For an arc between $\vartheta_0$ and $\vartheta_0+\Delta\vartheta$
\beq
\cF(\varphi) =
 \exp\big[-{\Delta{\vartheta}}\fa\,\frac{\sigma_3}{2i}\big]\,, 
\label{phirot}
\eeq
 is  a \emph{rotation by angle ${\Delta{\vartheta}}\,{\fa}$ 
 around the 3rd axis} in isospace. 
 For a full circle we recover the Wu-Yang factor, $\cF(2\pi)=\fF$ in \eqref{flux+factor}.
The non-Abelian Wu-Yang factor  
\eqref{NAWYfactor} can be presented in terms of its Abelian counterparts  \eqref{ABflux}, 
\beq
\cF^{NA}\equiv{\fF} = 
\barray{lr}({\fF}^{AB}_{(\alpha/2)} &0\\
0 &({\fF}^{AB}_{(-\alpha/2)})\earray\,.
\label{FNAFAB}
\eeq
Intuitively~: ``the number of components doubled but the exponents got halved, $2\alpha \to \fa$'' when compared to \eqref{ABflux}.

The \NAPF $\fF$ 
 transforms covariantly under a gauge transformation,
${\fF}(x_0)\to g^{-1}(x_0)\,{\fF}(x_0)\,g(x_0)$\,,
 and conversely~: two curl-free vector potentials $A^{(1)}$ and $A^{(2)}$ can be shown to be gauge-equivalent if and only if  their Wu-Yang factors are conjugate \cite{NABAPRD, Asorey},
\beq
\fF^{(1)} = h^{-1} \fF^{(2)}\, h
\label{AsoreyTHM}
\eeq 
for some fixed $h\in\SU(2)$. In conclusion,  we confirm~:

\begin{thm}
The gauge-equivalent flat YM potentials are in 1-1 correspondance with the Wu-Yang factors up to conjugation by some fixed element of the gauge group.
 \label{flatYMclass}   
\end{thm}

\goodbreak
\section{(Pseudo)classical isospin}\label{Isosec}

The additional concept we needed is that of  
\emph{classical isospin}, represented by a 3-vector $\vec{Q}$ or alternatively, by an $\su(2)$ matrix, eqn. \eqref{Qvecmatrix}.
Limiting ourselves at a non-relativistic study
in flat space, the equations of motion of a classical isospin-carying particle put forward by  Kerner \cite{Kerner68}, and by Wong \cite{Wong70}, and further discussed in \cite{Trautman70,Cho75,Witten81,Balachandran76,Balachandran77,Sternberg77,Sternberg78,DuvalAix79,Duval78,DH82,Wipf85,Feher86} are, 
\besub
\begin{align}
\ddot{x}_\mu =& - Q^a{\,}F_{\mu\nu}^a\dot{x}^\nu\,,
\label{Kernereq}
\\[4pt]
D_t Q \,\equiv& \;\;
 \dot{Q} + [A_\nu \dot{x}^\nu,Q] = 0\,,
\label{isospineq}
\end{align}
\label{KWeqns}
\esub
where the ``dot'' means $ d/{dt}$ with
 $t$ the non-relativistic time.
 The first of these  is a generalized Lorentz-force equation, and the second says that the isospin is parallel transported along the space-time trajectory.
The latter  is gauge invariant whereas that of the isospin transforms covariantly when gauge transformations are implemented in the adjoint representation, 
\beq
{Q} \to  g^{-1} {Q} \, g\,
\Rarrow
D_{t}{Q}  \to g^{-1} D_{t}{Q}\, g\,.
\label{gaugeonIso}
\eeq 
 Moreover, deriving $|Q|^2 = Q^a Q^a$  and using the isospin equation \eqref{isospineq} implies that the \emph{
length of the isospin is a constant  of the motion},
\beq
|Q|^2 = |Q_0|^2 = \const\,,
\label{Qlength}
\eeq
where $Q_0$ is an arbitrary point in an adjoint orbit 
$\cO=\big\{g^{-1} Q_0 g \,| \,g\in G\big\}$ of the gauge group \cite{Sternberg77,Sternberg78,DuvalAix79,Duval78,DH82,Wipf85,Feher86,HPAAix79}. For $G=\SU(2)$ the orbit is a 2-sphere $\IS^2$ of radius $|Q_0|$ in isospace.  

\subsection{Internal charges and symmetries 
 }\label{IntSymSec}

As emphasised by Yang and Mills and recalled in the Introduction, the distinction between protons and neutrons is based on their charges. The requirements  
are satisfied  by choosing, in each point, a
``direction field'' in the Lie algebra $\fG=\su(2)$, represented by vector $\hPsi$ of unit length,
$|\hPsi|=1$,  which should behave under a gauge transformation as a (normalized) \emph{adjoint Higgs field},
\beq
\hPsi \to g^{-1}\hPsi g\,.
\label{Psicov}
\eeq
Once  a choice has been made,
 the electric charge of our particle will be defined as the \emph{projection of the isospin} onto $\hPsi$,
\beq
\medbox{
 q = -2\,\tr(Q\cdot\hPsi) = Q^a\,{\hPsi}^a\,.
 }
\label{qdef}
\eeq
Its gauge invariance follows from combing \eqref{Psicov} with that of the isospin vector, \eqref{gaugeonIso}. 
But is it dynamically conserved ?
Before studying the \NABA\!, let's first consider a different but similar problem, namely that of motion in the field of a Wu-Yang monopole \cite{WuYang69},  
\beq
A_i^{a}=\epsilon_{iak}\frac{x_k}{r^2}\,. 
\qquad
\label{WYmonopA}
\eeq  
This  ``hedgehog'' Ansatz
identifies the internal isospace, $\su(2)$, 
 with the external $3$-space. Then projecting the isospin onto the radial direction, 
\beq
\hPsi^a=\frac{x_a}{r}\,.
\label{hedgehogHiggs} 
\eeq
The auxiliary direction field $\hx=\frac{x_a}{r}$ behaves as an adjoint Higgs field in the particular ``hedgehog'' gauge and then \eqref{qdef} allows us to recover the charge,
\beq
q = Q^a \frac{x_a}{r}\,,
\label{WYcharge}
\eeq
whose conservation follows from
$ D_\mu\hPsi=0\,.$ 

The charge \eqref{qdef} generalized to \NA monopoles is conserved  outside the monopole core, where $\hPsi$ is the direction of the Higgs field, which for large $r$ becomes covariantly constant   
 \cite{tHooft,Polyakov,GoddardOlive}.
 
 More generally, when a covariantly constant ``direction field" $\hPsi$ does exist then we posit, consistently with what we said in the introduction~: 

\begin{definition}
The particle is a proton if its internal charge \eqref{qdef} is $1$, and it is a neutron if its charge is $0$.
\end{definition}

The charge-based proton/neutron distinction 
 makes us  wonder: when do we get an \emph{ordinarily}, and not only covariantly, $D_t(\,\cdot\,)=0$, conserved charge, 
 \beq
 \dot{q}= 0\, ?
 \label{qdot0}
 \eeq  
The parallel-transport equation \eqref{isospineq} implies at once that this is so when $\hPsi$ is convariantly constant,
\beq
D_\mu\hPsi = \p_\mu \hPsi + [A_{\mu} , \hPsi] = 0\,. 
\label{convconstdir}
\eeq

However diatomic molecules  \cite{MSW,Jackiw86,KernerFest} 
provide us with an example where no covariantly constant direction fields and thus no conserved charges.
Our investigations in sec.\ref{NABAIntSymSec} then show that this is also what happens in the \NABA --- with the remarkable exception when \emph{the \NA flux is quantized} as in  \eqref{NAfluxquant}. 

\subsection{Conserved quantities}\label{CCharge}

Conserved quantities are usually derived from space-time symmetries by Noether's theorem \cite{JackiwManton,AbbottDeser}. Below we argue that the  charge \eqref{qdef} is generated by an  \emph{internal symmetry} \cite{HR-PRD,HR-color}.
We first recall some general facts 
\cite{Schwarz77,Harnad79,ForgacsManton,JackiwManton,Jackiw80,DH82,Wipf85,Feher86}.
A vectorfield $X$ is a symmetry for a gauge field if the variation  of the vector potential can be 
compensated by a gauge transformation, 
\beq
L_XA = DW \quad\text{\small or equivalently}\quad
F(X\,, \cdot\,) = D\psi
\where \psi = W - A(X)\,
\label{Fgensym}
\eeq
 for some Lie-algebra valued fields  $W$ or $\psi$, respectively
\cite{Schwarz77,ForgacsManton,Harnad79,Jackiw80}. 
Then the adapted form of the Noether theorem allows one to derive conservation laws \cite{JackiwManton,Jackiw80,DH82}.

As a first illustration, we consider the angular momentum in the field of a Wu-Yang monopole, \eqref{WYmonopA}, for which the general formul{\ae} yield the conserved  total angular momentum \cite{vHolten,HP-NGOME,KernerFest},
\beq
\vJ=\vx\times\vpi-q\hx
\label{WYJ}
\eeq
where $q$ is the conserved charge \eqref{WYcharge}.
This expression is
 analogous to the one in the field of a Dirac monopole \cite{JackiwRebbi76,Hasi,JackiwManton,Jackiw80, DH82,Wipf85,Feher86,Jackiw86,vHolten,HP-NGOME}. 
Eliminating $\vpi$ in favor of $m\dot{\vx}=\vp=\vpi+\vA$, \eqref{WYJ} $\vx\times\vA$ and  $q\hx$ combine into the isospin vector, and the angular momentum is 
  separated into the sum of  orbital and internal terms,
\beq 
\vJ= 
\vL -\vQ\,, \where \vL =\vr\times \vp\,,
\label{WYangmom}
\eeq
highlighting the ``spin from isospin'' phenomenon \cite{JackiwRebbi76,Hasi,JackiwManton,Jackiw80,Chen:2008ag}. 
Similar formulae will show up again in sec.\ref{Spinfromisospin}. 


Now we spell out the general theory for internal symmetries and charges in the \NA \AB context.
We  first consider a general compact and connected  gauge group $G$ before returning to $G=\SU(2)$.

The symmetry condition \eqref{Fgensym} \emph{assumes} that the (infinitesimal) transformation is implemented on the fields by   the Lie derivative by the vector field $X$ on the spacetime.  
 But a gauge transformation acts on space-time trivially, $X=0$; Then \emph{how does it act} on  fields ? And if it has been implemented, is it a symmetry ? 
 The answer, which generalizes what we said in sec.\ref{IntSymSec} for the electric charge, proceeds in three steps  \cite{HR-PRD,HR-color}~:

\begin{enumerate}
\item
Question~: \textit{Which elements of the gauge group can be \underline{implemented} on the gauge field, \ie, for which the action of a $g\in G$  on the gauge potential can 
 be defined ?}

\item
Question~: \textit{Let $K$ be an implementable subgroup of the gauge group $G$  and $k \in K$. Does it act as a \underline{symmetry} ?}

\item 
Question~: \textit{What is the conserved quantity associated with  a
vector $\kappa$ of the Lie algebra $\fK$ of the internal symmetry group $K$ ?}

\end{enumerate}

Coming to details, 

\begin{itemize}
\item
A  subgroup $K$ of the gauge group $G$ is said \emph{implementable} if there exists, in each point $x$ of spacetime, an automorphism of $K$ into of the gauge group $G$, 
\beq
\Omega_{\{\cdot\}}(x) : K \to G \st
\Omega_{k_1k_2}(x)=\Omega_{k_1}(x)\Omega_{k_2}(x)\,.
\label{impkdef}
\eeq 

Such a mapping can be made to act on the gauge potential according to
\footnote{For comparison : a gauge transformation by $k=k(x)$ would act instead as $A_\mu \to k^{-1} A_\mu \, k + k^{-1} \partial_\mu k\,$. In bundle language, the gauge group acts from the right, while the implementation \eqref{kimpA} acts from the left \cite{HR-color}.},
\beq
k\cdot A_{\mu}(x) = 
{\Omega_k}(x)A_{\mu}({\Omega_k}(x))^{-1}-
\p_{\mu}\Omega_{k}(x)({\Omega_k}(x))^{-1}\,.
\label{kimpA}
\eeq 

If the underlying space has non-trivial topology, a  topological obstruction against implementing certain elements of the gauge group can arise, and $K$ will be a proper subgroup of $G$. This is what happens  for non-Abelian  monopoles \cite{Schwarz82,NelsonMano83,MarmoBala82,NelsonColeman84,HR-PRD,HR-color}. 
For the \NABA  no such obstruction does arise, though \cite{HR-PRD,HR-color}, therefore the general theory will be skipped; the interested reader is advised to check the literature.
However, a (sub)group $K$ can have \emph{different inequivalent} implementations, -- and it is what happens for the \NABA as it will be shown in sec.\ref{NABAIntSymSec}.
 
Returning to the general theory, let us assume that $K \subset G$ is an implementable subgroup. The infinitesimal action of a vector $\kappa \in \fk$, the Lie algebra of $K$, is then a  $\fG$-valued Higgs-type field $\homega_{\kappa}(x)$ normalized as $|\homega_{\kappa}|=1$
which acts on the gauge potential by covariant derivative
\cite{Schwarz77,ForgacsManton,Harnad79,JackiwManton,Jackiw80,HR-PRD,HR-color}, 
\beq
\homega_\kappa\cdot A_\mu = D_\mu \homega_{\kappa}\,.
\label{infimp}
\eeq 
Then the answers to the questions above are:

\item
$\kappa$ is an \emph{infinitesimal symmetry} when $\omega_{\kappa}$ is covariantly constant
\beq
\medbox{
D_\mu \homega_{\kappa} = 0\,.
}
\label{ocovconst}
\eeq

\item
The Noether theorem applied to an internal symmetry then says that
\beq
\medbox{
 q_{\kappa} = -2\,\tr(Q\cdot{\homega}_{\kappa}) = 
 Q^a\,{\homega}_{\kappa}^a 
}
\label{omegacharge}
\eeq
is ordinarily conserved, $\dot{q}_{\kappa}=0$ generalizing \eqref{qdef}. 
\end{itemize}

\smallskip
For {\sl aficionados} of fiber bundles and differential geometry \cite{Kobayashi,Husemoller,HR-PRD,HR-color}, we first recall two definitions: 

(i) the centralizer of a subgroup $K \subset G$ is defined as all $g$ elements of $G$ which commute with all $k\in K$,
\ie, whose adjoint action leaves all $k\in K$ invariant,
\beq
Z_G(K)= \Big\{g^{-1}kg=k\; | \; \text{\small for\ all}\;
k\in K\Big\}\,.
\label{centralizer}
\eeq 
The centralizer of $K = G$ itself is thus the center, $Z(G)$, of $G$.

(ii)  Using fiber bundle language \cite{Kobayashi,Husemoller}, the horizontal lift,
obtained by calculating the integral \eqref{WYLoop} of the YM potential along a loop, ends in the same fiber it started from, and can therefore be
obtained by (right) action of a group element.
The holonomy group consists of all possible values of the Wu-Yang factor \eqref{WYLoop}. 

In the \NA \AB context, the holonomy group consists of integer powers of the Wu-Yang factor \eqref{WYLoop}, calculated for an arbitrary loop which generates $\pi_1 \approx \IZ$ of the punctured plane.
Then we quote PROP. 4.3 of \cite{HR-color}~: 

\begin{thm}
A subgroup $K$ of the gauge group $G$ is an internal symmetry group for the YM field $A=A_{\mu} dx^{\mu}$ if and only if the centralizer of $K$ in $G$, $H=Z_G(K)$, contains the holonomy group of $A$. The bundle then reduces to one with structure group $H=Z_G(K)$ obtained by solving the internal symmetry equation \eqref{ocovconst}.
The entire $G$ is itself an internal symmetry if and only if the holonomy group belongs to the center, $Z(G)$.
\label{holonomy}
\end{thm}

In  sec.\ref{NABAIntSymSec} we show that,
in the \NA \AB context, the symmetry condition \eqref{ocovconst} may have several inequivalent solutions. 

\section{Isospin precession}\label{IsoprecSec}

As emphasised in the introduction, 
\emph{there is no classical \AB effect} in the Abelian theory~: charged particles move freely in the curl-free magnetic potential. 
The effect is purely \emph{quantum mechanical} and is related to the phase freedom  \cite{AharonovBohm,Chambers, Mollen,Tonomura87,Peshkin89,Tonomura89,
HPAAix79}.

What about the non-Abelian case ?
This is the question to which this paper is devoted. 

The isospin couples to the \NA field \cite{Kerner68,Wong70} and the results obtained in an analogous but different context, namely for diatomic molecules \cite{MSW,Jackiw86} hint at \emph{isospin precession}, recovered 
 also at the (pseudo-) classical level \cite{KernerFest}.
  
In this paper we find (pseudo-) classical isospin precession also in the \NA \AB context.
\goodbreak

\begin{figure}[h]
\includegraphics[scale=.55]{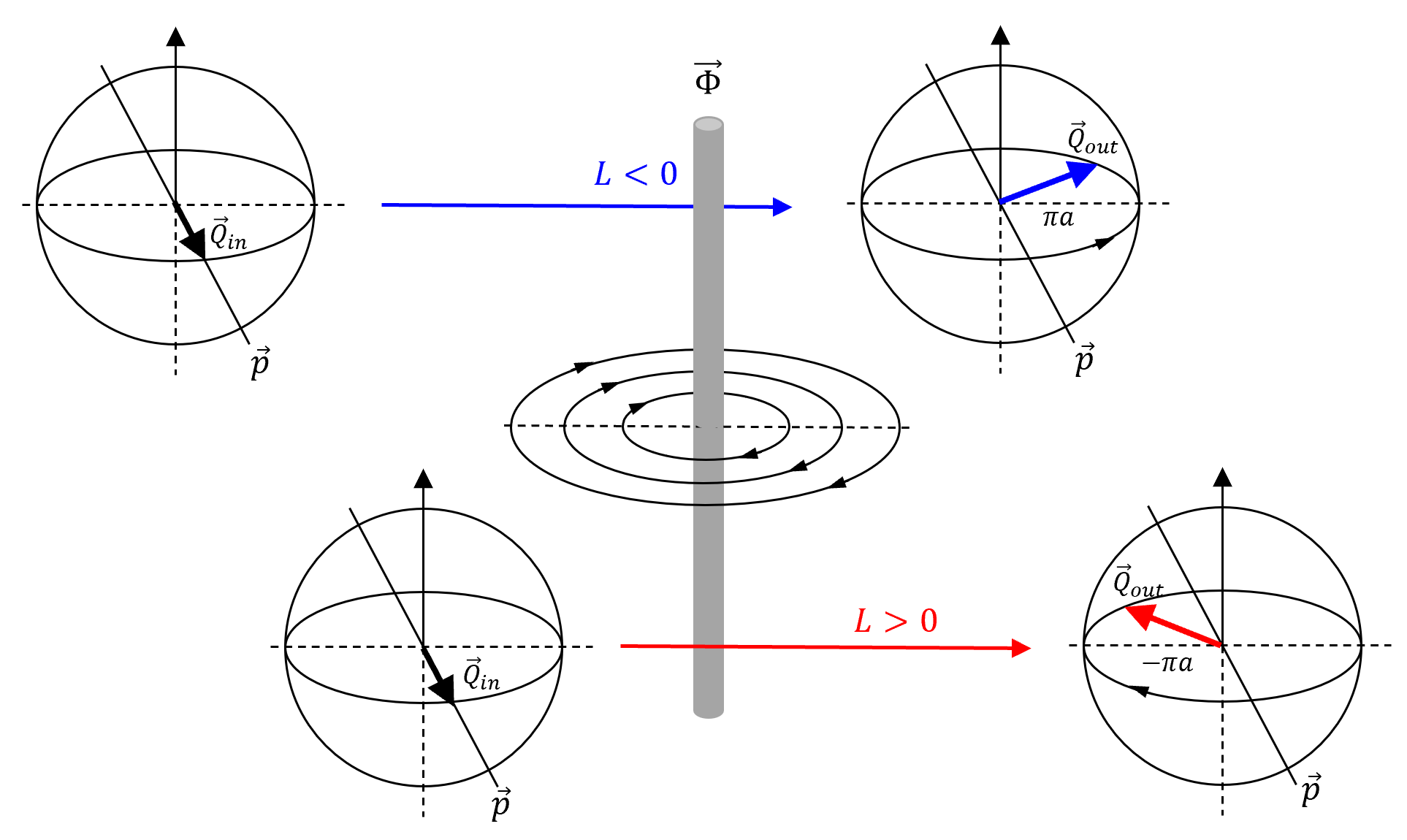}
\vskip-4mm
\caption{\textit{\small Scattering of a nucleon on a non-Abelian flux $\vec{\Phi}$, whose direction is perpendicular to the proton/neutron axis. The isopin is rotated by an angle {${\bf {\pm} \Delta\vartheta\,{\fa}}$}  with the sign depending on the side the particles passes, \ie, the sign, \red{${\bf {L > 0}}$} or \blue{${\bf {L < 0}}$}, of the (3rd component) of the orbital angular momentum.
The rotation of the isospin is generated by the \NAPF $\cF(\Delta\vartheta)$ in \eqref{phirot}. } 
\label{NABAproton}
}
\end{figure}

 Focusing our attention at the gauge group $G=\SU(2)$ we consider a non-Abelian flux confined to a thin cylinder
 upon which particles with isospin are scattered, as illustrated schematically in FIG.\ref{ABsetup}  
 and, in more detail, in FIG.\ref{NABAproton}. 

We shall need a first but important result. Unlike as for monopoles \cite{JackiwRebbi76,Hasi,JackiwManton,Jackiw80,vHolten,HP-NGOME}, the \emph{orbital angular momentum alone,}
\beq
\vL = \vx\times\dot{\vx}\,,
\label{Lorbital}
\eeq
(for unit mass)
\emph{is} conserved, for \emph{all values of the \NA flux}, in the \NA \AB context. This is seen
directly by using \eqref{Kernereq} with $F \equiv 0$:
$
\dot{\vL} = \dot{\vx}\times\dot{\vx}+\vx\times\ddot{\vx}
=0.\,
$
The sign of ${\vL}$ will be used to distinguishes the half-planes
in the two sides of the flux, as illustrated schematically in FIG.\ref{NABAproton}. Further aspects of 
the angular momentum will be studied in sec. \ref{Spinfromisospin}.
  
The classical dynamics is given by the Kerner-Wong equations
 \eqref{KWeqns}. Outside the cylinder the Yang-Mills field is  flat,\, $F = DA=0$, therefore the generalized Lorentz force  in \eqref{Kernereq} vanishes and  \emph{the space-time motion is free}, just like in the Abelian case.
\goodbreak 

As to the isospin, 
we first  remark that, in the free case, $a=0$ in the diagonal gauge \eqref{diagpot}, therefore the flux 
$\vPhi$  vanishes  and the isospin eqn. \eqref{isospineq} implies that,
\beq
Q(t) = Q_0 =\const.
\label{noQmot}
\eeq
However switching to another
gauge \eqref{gaugeonIso} by an \emph{arbitrary} gauge transformation $g(x^{\mu})$,
\beq
Q \to \tilde{Q}(t) = g^{-1}\big(x^\mu(t)\big)\, \tilde{Q}_0\,g\big(x^\mu(t)\big)\,,
\label{gQmot}
\eeq
the trajectory will look substantially different
--- but this is a mere  gauge artifact.

Do we find genuine, non-gauge-artifact classical motions~?

 When
the  \NA flux $\Phi$ and the incoming isospin $\vec{Q}_{in}$ are parallel in internal isospace (considered by Wu and Yang) the answer is negative by \eqref{isospineq}. 
However in the second case proposed in \cite{WuYang75}, \ie, when  $\vQ_{in}$ and the non-Abelian flux $\vPhi=(\Phi^a)$ are \emph{not parallel} in isospace, then genuine \emph{internal motion} can be found, as we show now. 

To explain how this comes about, let us
 assume for simplicity that $\vec{Q}_{in}$ is perpendicular to the flux in isospace, $Q_{in}^3=0$, as depicted in FIG. \ref{NABAproton}. 
 
First of all, we fix our terminology: consistently with the Yang-Mills argument \cite{YangMills} quoted in the introduction we shall call, in the diagonal gauge, the  particle \emph{incoming} from far on the right,
\beq
\left\{\barraynb{clllll}
\text{proton}\quad &\text{if}\qquad
&Q_{in}^p&=&\smallover{1}/{2i}\sigma_1
&=\frac{1}{2i}{\tiny \barray{cc}0 &1 \\ 1 &0\earray}\,
\\[12pt]
\text{neutron}\quad &\text{if}\qquad
&Q_{in}^n&=& \smallover{1}/{2i}\sigma_2
& = \smallover{1}/{2}{\tiny \barray{cc}0 &-1 \\ 1 &0\earray}
 \earraynb\right.\;
\label{Qinpn}
\eeq
which both have unit length and are orthogonal to each other w.r.t. the scalar product \eqref{omegacharge},   
\beq
-2 \,\tr(Q_{in}^2)=1
\aand
\tr(Q^n_{in}\cdot Q^p_{in})=0\, .
\label{orthonorm}
\eeq
This fixes the choice at one point -- and, as  Yang and Mills say \cite{YangMills}, at all other points by gauge covariance.  
\goodbreak
 
\begin{figure}[h]
\includegraphics[scale=.5]{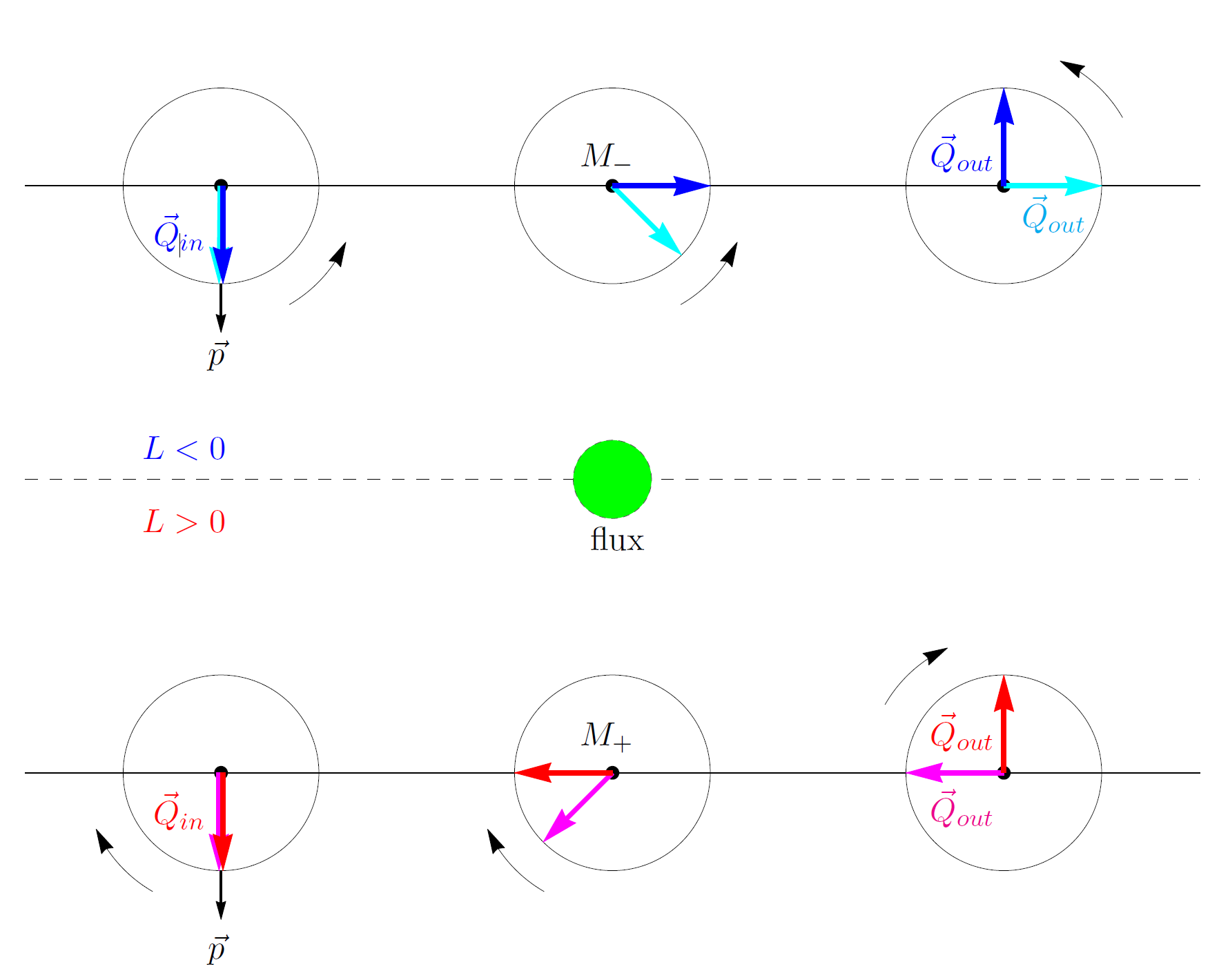}
\\
\vskip-4mm
\caption{
 \textit{\small The ``isospin clockwork"  for \NA flux $\Phi_a=-i{\pi}a\sigma_3$, ($a\neq0$).  
The incoming proton (at $L_\pm$) is aligned with ${\vec{p} = \sigma_1/2i}$ (at ``6 o'clock''),
${\bf Q^p_{in}}$= ${\vp}$.
For \red{${\bf L > 0}$} the isospin  rotates clockwise, but
for \blue{${\bf L < 0}$} it  rotates counter-clockwise. 
For odd integer  flux  (${\bf a = 1}$, for example),
drawn in \red{\bf red} and \blue{\bf blue},
$\vec{Q}_{out}^{p} = -\vec{Q}_{in}^{p}$ ($R_{\pm}$ at ``12 o'clock'') on both sides. However when $a$ is not an integer e.g.   
for ${\bf a=\half}$ (mod even integers),  
drawn in \magenta{\bf magenta} and \cyan{\bf cyan}, the isospin is lagged behind.
\magenta{$\vec{Q}_{out}^{p}$} and (``9 o'clock'') and \cyan{$\vec{Q}_{out}^{p}$} (``3 o'clock'') are both orthogonal to $\vec{Q}_{in}^{p}$  and are oppositely oriented in the two $L$-sectors. Similar behavior could be observed for any $\fa$.
\label{clockwork}
}}
\end{figure}

Now we study the motion of the isospin. Recall first that
 \eqref{Qlength} implies that $|Q|=\const$,  
therefore $Q$ moves on a circle. 
 Eliminating  $t$ in favor of the polar angle $\vartheta$, $\dot{\vartheta}$ drops out from the parallel transport equation \eqref{isospineq}, leaving us  with
$
\frac{dQ}{d\vartheta} + \big[A_{\vartheta},Q\big]=0\,. 
$
It follows that $Q^3=\const$. We put $Q^3=0$ for simplicity. Let us  assume  that the incoming particle is a proton, $Q_{in}=Q_{in}^p$. First consider $L >0$. 
Then 
\beq
\bigbox{
{Q}^p_{+}(\varphi) 
=\cos(\varphi{\fa})\,\frac{\sigma_1}{{2}i}-
\sin(\varphi{\fa})\,\frac{\sigma_2}{{2}i}\,
\quad\ie,\quad
{Q}^p_{+}(\varphi) 
=
\frac{1}{{2}i}\barray{cc}
0 &e^{+i\varphi{\fa}}
\\
e^{-i\varphi{\fa}}&0\earray\,, 
}
\label{P+inout}
\eeq
where we shifted to $\varphi=\vartheta-\pi$,
 $0 \leq \varphi  \leq \pi$.
The initial position, far on the left, has  $\varphi_{in} \approx 0$. 
A similar formula works for $L<0$ with
$\varphi=\pi-\vartheta$ which changes just a sign, 
\beq
\bigbox{
{Q}^p_{-}(\varphi) 
=\cos(\varphi{\fa})\,\frac{\sigma_1}{{2}i}+
\sin(\varphi{\fa})\,\frac{\sigma_2}{{2}i}\,,
\quad\ie,\quad
{Q}^p_{-}(\varphi) 
=
\frac{1}{{2}i}\barray{cc}
0 &e^{-i\varphi{\fa}}
\\
e^{+i\varphi{\fa}}&0\earray\,.
}
\label{P-inout}
\eeq
The motion of the isospin can be viewed as a rotation of the isospin around the 3rd axis,
\beq
\vec{Q}_{out}^p = \left\{\barraynb{lll}
R_{+\varphi{\fa}}\,\vec{Q}_{in}^p  &\text{\small for} & L \geq 0
\\[7pt]
R_{-\varphi{\fa}}\,\vec{Q}_{in}^p  &\text{\small for} & L \leq 0
\earraynb\right.\;,
\label{isorot}
\eeq
as suggested by Kerner \cite{Kerner68}, -- but the rotation is in opposite directions as shown in FIGs.\ref{NABAproton}-\ref{clockwork}.

The intuitive picture of the opposite shifts suffered on the two sides is~: sailing downwind resp. upwind shifts the phase by the relative phase difference.  

Notice however that for $\varphi>0$ the upper and lower values of the isospin in \eqref{P+inout} and
\eqref{P-inout}, respectively, \emph{do not match unless $\fa$ is an integer},
\beq
\fa = 2k\; \oor \;\fa = 1+2k
\label{quantaval}\,. 
\eeq

Similarly for an incoming neutron,  $Q^n_{in} =\smallover{\sigma_2}/{2i}$, we get,
\beq
{Q}^n_{\pm}(\varphi) 
=\cos(\varphi{\fa})\,\frac{\sigma_2}{{2}i}{\mp}
\sin(\varphi{\fa})\,\frac{\sigma_1}{{2}i}\,
\quad\ie,\quad
{Q}^n_{\pm}(\varphi) 
=
\frac{1}{2}\barray{cc}
0 &-e^{\pm{i}\varphi{\fa}}
\\
e^{\mp{i}\varphi{\fa}}&0\earray\,.
\label{N+-inout}
\eeq
The formulae \eqref{P+inout}-\eqref{P-inout} and \eqref{N+-inout} will be recovered  
 in sec.\ref{IsoprecSec}, \# \eqref{PMQ} using a different framework.
 
\begin{figure}[h]
\includegraphics[scale=.6]{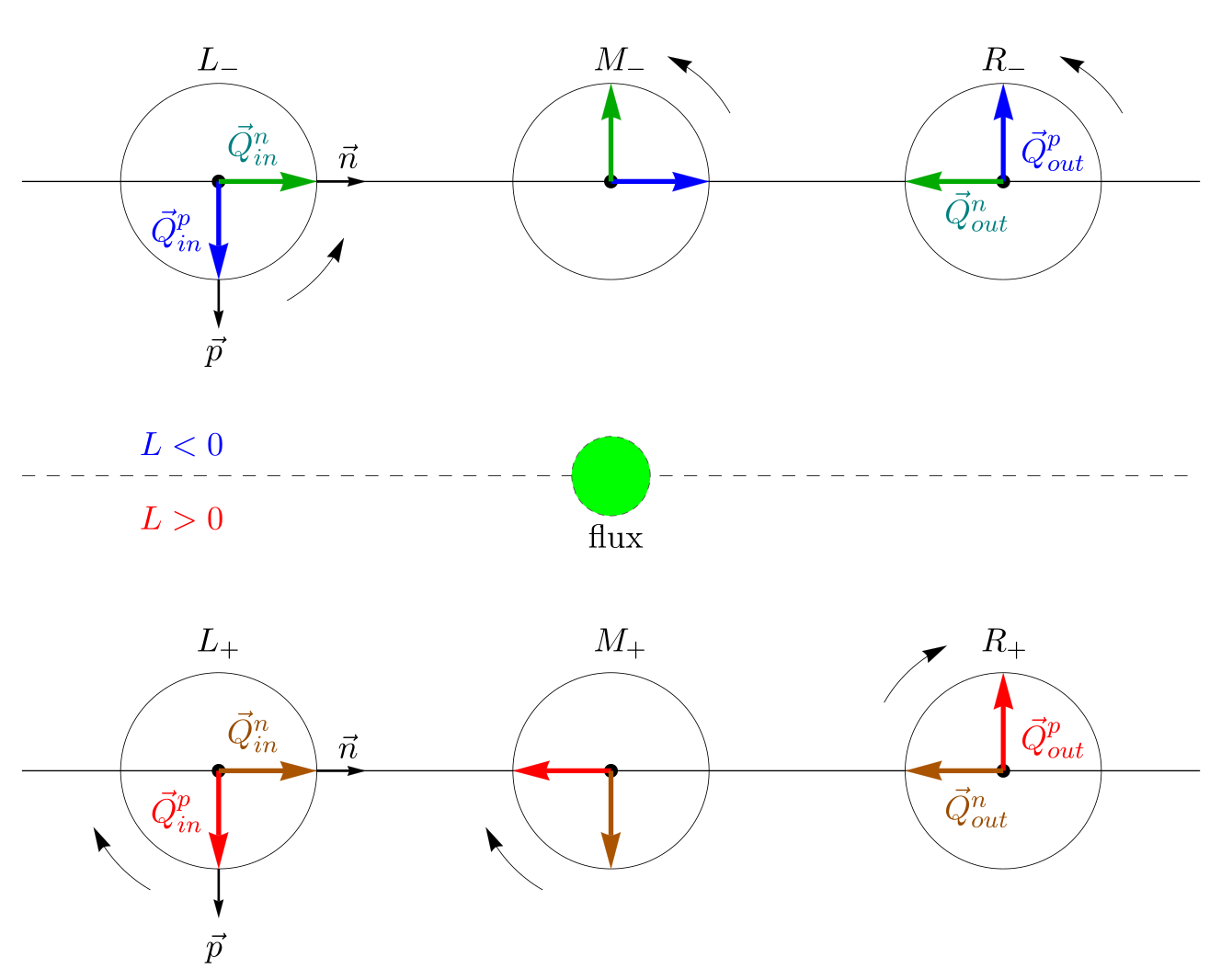} 
\\
\vskip-4mm
\caption{\textit{\small The ``isospin clockwork" for  \textbf{protons} and \textbf{neutrons} for $a=1+2k$  where $k$ is an integer.     
 \ie, for flux $\Phi=-i(1+2k)\pi\sigma_3$. The incoming proton 
 (in \red{red} and \blue{blue} is aligned with $\vec{p}$, $Q_{in}^p= \sigma_1/2i$ (``6 o'clock'') as in FIG.\ref{clockwork}. 
The incoming neutron (in \purple{purple} and \dgreen{green}) is along $\vn$  (at ``3 o'clock''), 
$Q_{in}^n =\bsigma_2/2i$ on both sides of the flux. 
For \red{${\bf L > 0}$} the isospin rotates clockwise for both $Q^p$ and $Q^n$ and for \blue{${\bf L < 0}$} they rotate counter-clockwise. $Q^p$ and $Q^n$ turn rigidly, remaining orthogonal to each other. 
$\vec{Q}_{out}^{p,n} = -\vec{Q}_{in}^{p,n}$ (``12 o'clock'' and ``9 o'clock", respectively) on both sides, consistently with $S=-\II$ in \eqref{II-II}.
The centers of the ``shifting clocks'' are at  
$\varphi = 0,\ \pi/2, \, \pi$.
}
\label{a=1PN}
}
\end{figure}

\section{Isospin scattering}\label{ScattSec}

We are be particularly interested by the value of the isospin  far on the right,  at ${\varphi}_{out}=\pi$, 
 shown in FIGs.\ref{NABAproton}-\ref{clockwork}-\ref{a=1PN}.
In other words, we want to study the \NABA  from the viewpoint of scattering theory \cite{NABA82}.
In terms of the polar angle, 
$\Delta\vartheta=\pm\pi$ ,
\beq
{Q}_{out}^p 
= 
\frac{1}{{2}i}\barray{cc}
0 &e^{\pm{i}\pi{\fa}}
\\[4pt] 
e^{\mp{i}\pi{\fa}}&0\earray\,,
\qquad
{Q}_{out}^n 
=
\frac{1}{2}\barray{cc}
0 &-e^{\pm{i}\pi{\fa}}
\\
e^{\mp{i}\pi{\fa}}&0\earray\,,
\label{QinoutC}
\eeq
with the sign depending on which side of the flux line had the particle passed, correlated in turn with the sign of 
 the vertical component, $L\equiv L_z$,  of the conserved 
 orbital the angular momentum $\vL = \bx\times \dot{\bx}$ in \eqref{Lorbital}. 
The rotations in \eqref{QinoutC} [or more generally in \eqref{P+inout}-\eqref{P-inout}] on the two sides do not match for  $L\to 0$ in general, though.
However in the non-trivial exceptional case ${\fa}=1+2k$, for example, the outgoing result becomes side-independent,
\beq
{Q}_{out}^p = 
- \frac{1}{{2}i}\sigma_1 = - Q_{in}^p
\,
\label{outa=1}
\eeq
consistently also with $R_{\pm\pi}= -\II$ in \eqref{isorot} and shown in FIG.\ref{a=1PN}. 
More generally, the isospin is reversed in both sectors.
 The r\^ole of the  \emph{minus} sign here will be explained in sec.\ref{IntSymSec}.
\goodbreak 

\begin{figure}
\includegraphics[scale=.45]{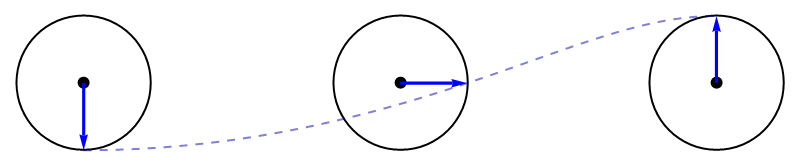}
\qquad\text{pL-}
\\
\includegraphics[scale=.45] {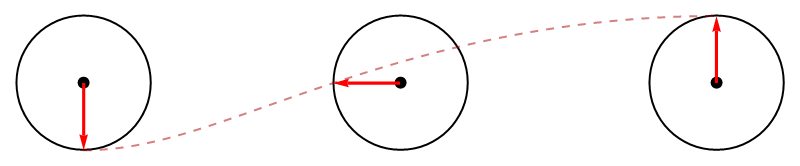} 
\qquad\text{pL+}
\\[8pt]
\includegraphics[scale=.45] {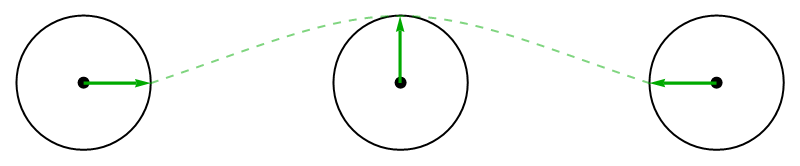}
\qquad\text{nL-}
\\
\includegraphics[scale=.45] {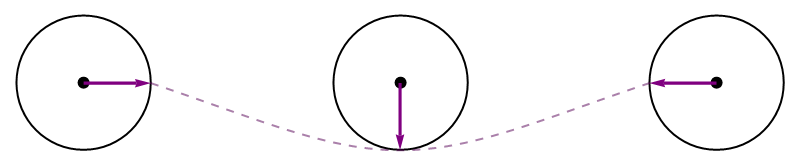}
\qquad\text{nL+}
\\ 
\caption{\textit{\small Isospin motion for ${\fa}=1$ for both protons \eqref{a=1Pcyclo} and neutrons \eqref{a=1Ncyclo}, in both respective half-planes. The trajectories follow cycloid segments, which all satisfy 
$\vQ_{out} = -\vQ_{in}$.}
\label{protoncyclo}
}
\end{figure}

Eqns. \eqref{P+inout}-\eqref{P-inout} with  $\Delta\theta=\pi$  allow us the express the result in terms of the Wu-Yang factor \eqref{flux+factor} in matrix resp. in vector form \cite{NABA82},
\beq
{Q}^p_{out} = \big(\fF\big)^{-1/2}{Q}^p_{in} \,\big(\fF\big)^{1/2}
\;\;\;\ie,\;\;\;
\vec{Q}^p_{out} = \fF\,\vec{Q}^p_{in}\,. 
\label{Qtraj}
\eeq
These formulae,  
consistent with \eqref{vecmatimp}, confirm that the isospin is  rotated by $\cF(\varphi)$ cf. \eqref{phirot}, as 
 shown in FIGs.\ref{clockwork} and \ref{a=1PN}. 
They are gauge-covariant when the incoming direction  also transforms covariantly,
$ 
Q_{out}^p  = \fF \,Q_{in}^p  \to   
(g^{-1} \fF g) (g^{-1} Q_{in}\, g)= g^{-1} (\fF\, Q_{in}) g = g^{-1} Q_{out}\,g\,.
$
For both  exceptional values $\fa=2k$ and $\fa=1+2k$ the 
phase difference accumulated  on the two sides is $\pm 2\pi$, implying that
the  \NA fluxes differ on  by an even integer and thus yield identical Wu-Yang factors, 
\beq
\fF= \left\{\barraynb{rlc}
\II &\when &{\fa}=2k \;
\\
-\II &\when &{\fa}=1+2k
\earraynb\right.\; .
\label{II-II}
\eeq 
 Therefore $\vec{Q}_{out}=\pm\vec{Q}_{in}$ on  both half-planes, \eqref{outa=1}, as shown  in FIGs. \ref{clockwork} and \ref{protoncyclo}.
\goodbreak
 
The \NIPF for the half-loop \eqref{phirot}
is the {square root} of the Wu-Yang factor \eqref{flux+factor},
\beq
\cF(\pm\pi) = \big({\fF}\big)^{\pm{1/2}} 
=
\barray{cc}e^{{\pm}i\pi(\half{\fa}+k)} 
& 0 \\ 0 & e^{{\mp}i\pi(\half{\fa}+k)}\earray\,,
\label{NAFAB}
\eeq 
where $k$ is an integer \footnote{$\cF$ in \eqref{NAFAB} acts on the $\su(2)$ matrix $Q$ in the adjoint representation. In vector terms, $Q \leadsto \vQ$, the arguments of the exponential are doubled, cf. \eqref{Qtraj}.}. 
This expression is explained as follows. The Wu-Yang factor is obtained by integrating along a generating loop of $\pi_1$ put together from two half-loops, starting from $\vartheta=-\pi$, going first in the $L > 0$ half-plane to the turning point at $\vartheta=0$ 
 and then completing the loop by returning  to $\vartheta=\pi$ in the $L < 0$  half-plane. Each half-loop contributes equally. Thus 
$ 
 \fF
 =\cF(+\pi)\cF^{-1}(-\pi) = 
 \diag\big(e^{{i}{\pi}{\fa}},e^{-{i}{\pi}{\fa}}\big)\,. 
$  
  
Hence we focus our attention at the odd case. 
 Identifying the external and internal coordinates by 
  mimicking monopole theory 
  \cite{WuYang69,tHooft,Polyakov,GoddardOlive}, we get a sort of ``\AB hedgehog''.
 Then the trajectory of the isospin is that of a point on a rolling circle whose center moves on the spatial trajectory, --- \ie, a \emph{cycloid segment}, depicted in FIG.\ref{protoncyclo}. In detail,

$\bullet$ For a proton
\besub
\begin{align}
\red{
Q_1^p = |\vQ_0|\big(\varphi - \sin\varphi\big),} 
\quad
&
\red{Q_2^p = -|\vQ_0|\cos\varphi,}
\quad
0\leq\varphi=\red{\vartheta-\pi} \leq \pi
&\red{\text{L} > 0}
\\[3pt]
\blue{Q_1^p = |\vQ_0|\big(\varphi + \sin\varphi\big),} 
\quad
&
\blue{Q_2^p = - |\vQ_0|\cos\varphi,}
\quad
0\leq\varphi=\blue{\pi-\vartheta} \leq \pi
&\blue{\text{L} < 0}
\end{align}
\label{a=1Pcyclo}
\esub
$\bullet$ For a neutron
\besub
\begin{align}
\purple{
Q_1^n = |\vQ_0|\big(\varphi + \cos\varphi\big),} 
\quad
&
\purple{Q_2^n = -|\vQ_0|\sin\varphi,}
\quad
0\leq\varphi=\purple{\vartheta-\pi} \leq \pi
&\purple{\text{L} > 0}
\\[3pt]
\dgreen{Q_1^n = |\vQ_0|\big(\varphi + \cos\varphi\big),} 
\quad
&
\dgreen{Q_1^n = \;\;\;|\vQ_0|\sin\varphi,}
\quad
0\leq\varphi=\dgreen{\pi-\vartheta} \leq \pi
&\dgreen{\text{L} < 0}
\end{align}
\label{a=1Ncyclo}
\esub

\kikezd{S-matrix} \cite{NABA82}. 

Eqn \eqref{Qtraj}  allows us to consider a pseudo-classical S-matrix defined, separately, on the two half-planes distinguished by the sign of the orbital angular momentum $L$ 
\beq
S_{\pm} = \cF(\pm\pi) = \barray{cc}e^{{\pm}i\pi(\half{\fa}+k)} 
& 0 \\ 0 & e^{{\mp}i\pi(\half{\fa}+k)}\earray\,
\label{Smatrix}
\eeq
cf. \eqref{NAFAB} and \eqref{isorot}. The $\pm$ expressions match for ${\fa}=0$ or ${\fa}=1$ mod. even integers.
 In conclusion, 
 \goodbreak

\begin{thm}
When the incoming isospin $\vec{Q}_{in}$ and the \NA flux $\vPhi$ are \underline{not} parallel,
 the isospin vector $\vec{Q}$ \underline{precesses}
 in opposite directions, depending 
 on which side the  particle goes and distinguished by 
  the sign of the orbital angular momentum $L$.
The Wu-Yang factor $\fF$  \eqref{WYLoop} acts as ``classical S-matrix'' by rotating the isospin,
\beq
{Q}_{out} = S\,{Q}_{in}= (\fF)^{-1/2}{Q}_{in} (\fF)^{1/2}
\label{clSmatrix1}
\eeq\vskip-5mm
\;\ie,
\beq
\vec{Q}_{out} =  S\,\vec{Q}_{in} \where S = R_{\pm{\pi}{\fa}},
\label{clSmatrix}
\eeq
illustrated in FIGs. \ref{clockwork} and \ref{a=1PN}.
In the odd case \ie, for \NA flux 
\beq
\Phi^{odd}= (1+2k)\,\Phi_0,\qquad \Phi_0=\pi\sigma_3/i\,,
\label{a1flux}
\eeq 
 the S-matrix is minus the identity on both sides,
cf. \eqref{II-II} and 
 FIG.\ref{a=1PN}.
\label{claQrot}
\end{thm}
\goodbreak

For non-integer values of ${\fa}$ the S-matrix will be side-dependent. For ${\fa}=1/2$ (shown in FIG.\ref{clockwork} in \magenta{magenta} and in \cyan{cyan}), for example,
\beq
{Q}_{out}^p ({\fa}=\half)
=  
\left\{\barraynb{clll}
 -\frac{\sigma_2}{{2}i} &\quad \text{\small for} &\quad
 L > 0
\\[6pt]
+\frac{\sigma_2}{{2}i} &\quad\text{\small for} &\quad
L < 0
\earraynb\right.
\label{Qahalf}
\eeq
so that the incoming the isospin is rotated by {$\mp\pi/2$}, depending on the side.

These properties are related to charge conservation, studied in the next section. 
\goodbreak

\section{Internal charges}\label{NABAIntSymSec}

Does the minus signs in \eqref{outa=1} imply that a proton would have become  an ``antiproton'' ?! We argue that \emph{no} ! The reason is \emph{gauge covarance}. Our clue is that following Yang-Mills \cite{YangMills} the proton/neutron identity of a particle is 
defined by its \emph{charge} derived  by  the Noether theorem, and  not by the apparent value of its isospin, which is gauge-dependent. The charge should be defined by setting up a direction field, as we now explain by
applying the general theory 
to the \NABA outlined in sec. \ref{IntSymSec}.

\subsection{Implementability and internal symmetry}\label{NABAimp}

\vskip-5mm
\kikezd{Implementability}. Unlike as for monopoles \cite{NelsonMano83,MarmoBala82,NelsonColeman84,HR-PRD,HR-color}, no topological obstruction arises against implementing gauge transformations in the \NA \AB case. Leaving the mathematical subtleties to the literature, we merely cite the statement proved in \cite{HR-PRD,HR-color,NABAPRD}~:
\begin{thm}
In the \NABA the entire gauge group $\SU(2)$ is implementable, namely as in \eqref{kimpA}.
However we have 
 \underline{two inequivalent implementations},
namely a trivial one and a twisted one, defined, in the diagonal gauge, by 
\beqa
\Omega_{k}^{(1)}(x)=k=\big(k_{ij}\big)
\;\aand\,\;
\Omega_{k}^{(2)}(x) =
\barray{cc} k_{11}& e^{i\varphi} k_{12} \\
e^{-i\varphi} \bar{k}_{12} & \bar{k}_{11}\earray\,,
\label{twogroupimps}
\eeqa %
where 
$\det k = |{k}_{11}|^2+ |k_{12}|^2 = 1$. 
Infinitesimally, 
\beq
\homega_{\kappa}^{(1)}(\varphi) = \; {\frac{1}{2i}}\kappa={\frac{1}{2i}}\big(\kappa_{ab}\big) \,
\aand\,
\homega_{\kappa}^{(2)}(\varphi) = {\frac{1}{2i}}\barray{cc}
\kappa_{11} &e^{i\varphi}\, \kappa_{12} 
\\[2pt]
e^{-i\varphi}\,\overline{\kappa}_{12} &-\kappa_{11}
\earray\,, 
\label{twoalgimps}
\eeq 
\!respectively, where we 
 normalized to $|\homega|=1$. 
 \label{impthm}
\end{thm}
\goodbreak

Recalling that  $\varphi=\pm(\vartheta-\pi)$, the second implementation can also be written  
 in terms of the polar coordinate 
 $0\leq \vartheta \leq2\pi$  as,
 \beq
\Omega_{k}^{(2)}(\vartheta) = \barray{cc}
k_{11} &-e^{{\pm}i\vartheta}\, k_{12} 
\\[2pt]
-e^{{\mp}i\vartheta}\,\overline{k}_{12} &-k_{11}
\earray,
\quad
\homega_{\kappa}^{(2)}(\vartheta) = {\frac{1}{2i}}\barray{cc}
\kappa_{11} &-e^{{\pm}i\vartheta}\, \kappa_{12} 
\\[2pt]
-e^{{\mp}i\vartheta}\,\overline{\kappa}_{12} &-\kappa_{11}
\earray,
\label{OPsitheta} 
 \eeq
 respectively,
with the sign depending that of the orbital angular momentum, $L$.
Reassuringly, for $\kappa=\sigma_1$ 
we recover an incoming proton, 
\beq
\homega^{(1,2)}(\varphi=0)= \homega^{(1,2)}(\vartheta=\pi)= Q^p_{in}\,.
\label{incomprot}
\eeq 
The 2nd implementation, by $\Omega_{h}^{(2)}$, could actually be brought also to position-independent form
by gauge transformations, defined separately in the two contractible domains,
\beq 
\left\{\barraynb{llll}
g_-=
\diag(e^{i\vartheta/2},e^{-i\vartheta/2})
&\quad\text{\small in}\quad
V_-\; &=&\;\;0\leq \vartheta \leq \pi
\\[8pt]
g_+=\diag(e^{-i\vartheta/2},e^{i\vartheta/2})
&\quad\text{\small in}\quad
V_+ &=&\;\; \pi \leq \vartheta \leq 2\pi
\earraynb\right.\;
\eeq  
\beq
\Omega_{k}^{(2)}(\vartheta)\to g_{\pm}\Omega_{k}^{(2)}(\vartheta)g_{\pm}^{-1}=
\barray{rr} k_{11}& -k_{12} \\
-\bar{k}_{12} & \bar{k}_{11}\earray\,.
\eeq
However the gauge transformations do not match on the negative $x$ axes,
$ 
g_+(\vartheta=\pi) = 
 -i\sigma_3,
\,  
g_-(\vartheta=\pi) =  
 i\sigma_3
\,.
$
``Dressing'' $\Omega_{k}^{(2)}$ to a constant introduces a non-trivial transition function, $-\II$  \footnote{The inequivalent implementations correspond to the center of the gauge group \cite{HR-color}. The trivial one  corresponds to $\II$ and the twisted one in \eqref{twogroupimps} corresponds to $-\II$. For $G=\SO(3)$ one has only $\II$.}.
In what follows, we consider  the infinitesimal version.
 \goodbreak


\kikezd{Internal symmetries}.

Which of the  implementations \eqref{twoalgimps} are symmetries ? 
We recall that (as said in sec.\ref{IntSymSec})~:  $\kappa=(\kappa_{ab})\in\fG$ generates an internal symmetry when
 $\homega_{\kappa}$ is covariantly constant, $D_{\mu}\homega_{\kappa}=0$, cf. \eqref{ocovconst}. Then putting the covariant derivates to zero and taking into account that ${\fa}$ and ${\fa}-2k$ with $k$ an arbitrary integer are equivalent,  \eqref{a2k}-\eqref{fluxdiff}, 
\besub
\begin{align}
D_{\mu}\homega_{\kappa}^{(1)}  
= \;&\; \;\;\; \;\,\frac{{\fa}-2{k}}{i}\;\;\;\;\;\;\quad
\barray{cc} 0 & \kappa_{12}
\\
 -\overline{\kappa}_{12} & 0
\earray \,,
\\[14pt]
D_{\mu}\homega_{\kappa}^{(2)} =
\; &\frac{{\fa}-(1+2k)}{i}\,\barray{cc}
0 &\,e^{i\varphi}\kappa_{12}
\\
-\,e^{-i\varphi}\overline{\kappa}_{12}  &0\earray\,. 
\label{Dtwistimp}
\end{align}
\label{twoDimps}
\esub 
Thus the covariant derivatives vanish and thus yield an internal symmetry only for 
\beq
\text{\small either}\quad {\fa} = 2k
\oor
{\fa} = 1+2k \quad (\text{$k$ \ an  integer})\,,
\label{intsym01}
\eeq 
respectively. In both cases, the \emph{enclosed magnetic flux is quantized} for both  series, 
\besub
\begin{align}
\Phi_k^{even} = &\;\;\;\; (2k)\,\,\Phi_0, \qquad\Phi_0=(-i\pi \,{\sigma_3}),\, \qquad k=0, \pm1, \pm2,\, \dots\,,
\label{evenflux}
\\[6pt]
\Phi_k^{odd} \; = &\; (1+2k)\,\Phi_0, \quad\Phi_0=(-i\pi\,  {\sigma_3}), \quad\:  k=0, \pm1, \pm2,\, \dots\, ,
\label{oddflux}
\end{align}
\label{NAfluxquant}
\esub 
Comparison with \eqref{ABflux} shows that the Abelian series splits into two series, one even and one odd.
The Wu-Yang factor  takes its values in the center $Z(G) = \{\II,-\II\}$ of the gauge group $\SU(2)$~\footnote{
For $\SO(3) = \SU(2)/\IZ_2$ the center would be just $\{\II\}$ and the twisted implementation would not be symmetry, leaving us  only with the trivial one.}\,, as seen in \eqref{II-II}. 
We have, of course  also
\beq
\homega_{\sigma_3} = \frac{\sigma_3}{2i}\,.
\label{vertomega}
\eeq

\begin{thm} For ${\fa}=0$ or for ${\fa}=1$ (mod even numbers) \ie, for Wu-Yang factor $\fF = \pm\II$,  each $\kappa\in\su(2)$  generates, 
  via the first resp. the second implementation in \eqref{twoalgimps}  an internal symmetry and thus generate $3$ conserved internal charges, \eqref{omegacharge}, 
\beq
\barraynb{lclc}
q_{b}^{(1)} =&\quad Q^{b} &\for &{\fa}=2k
\\[6pt]
q_{b}^{(2)} =& 
\;- 2\tr\left(Q\cdot{\homega^{(2)}}_{\sigma_b}\right)
&\for &{\fa}=1+2k
\earraynb
\label{NABAcharges}
\eeq  
where $b= 1,2, 3$ are internal indices. 
For non-integer values of ${\fa}$ in the vector potential \ie, for  $\fF\neq \pm\II$, there are no internal symmetries and thus no conserved internal charges. 
\label{3intcharges}
\end{thm}

Thus the conserved charges are $q_b^{(1)}= Q^{b}$ for the trivial implementation ${\fa}=2k$ and $q_b^{(2)}$  in the twisted case with vector potential ${\fa}=1+2k$. Then for the initial choices \eqref{Qinpn} we recover the proton, for $b=1$ and the neutron for $b=2$ with charges  $q_{proton} = 1$ and $q_{neutron} = 0$, respectively, 
consistently with \eqref{orthonorm}.
The conservation of the charges \eqref{NABAcharges}
follows also from covariant-constancy,
$
\dot{q}=-2\tr\big(D_tQ\cdot\hPsi+Q\cdot{D_t\hPsi}\big)=0.
$

The full gauge group $G = \SU(2)$ is a symmetry in both cases \cite{HR-PRD,HR-color}. 
However in the even case $a=2k$ 
the field \eqref{diagpot} is a pure gauge
\beq
A_{\mu}=g^{-1}\p_{\mu}g
\for 
g(x)= e^{ik\vartheta\sigma_3} \,
= \diag\big(e^{ik\vartheta} , e^{-ik\vartheta}\big)\,.
\label{Apuregauge}
\eeq
 
In the non-trivial odd case,  ${\fa}=1+2k$,  when the \emph{flux is quantized in {odd} units} as in \eqref{oddflux}. 
For this value of ${\fa}$ the field is \emph{not} a gauge transform of the vacuum~: although the  it   
 could seemingly be removed by gauge transforming by $g(\vartheta) = e^{i\vartheta\sigma_3/2}$ However  this function is {not} well-defined on the plane,  because $g(\vartheta=0)=\II$ but
 $g(\vartheta=2\pi)=e^{i\pi\sigma_3}=-\II\neq\II$ \footnote{
Remember that for ${\fa}=1+2k$ the S-matrix is minus the identity, 
$ S = -\II \,,$ 
as noted in \eqref{II-II}.}.

\subsection{Conserved charges in the non-Abelian Aharonov-Bohm effect
 }\label{NABAcharge}

Henceforth we focus our attention at ${\fa}=1+2k$ an odd integer, we  fix  the implemention to be that of the initial proton direction,  \eqref{Qinpn}, 
$
Q_{in}^p=\smallover{1}/{2i}\sigma_1
=\frac{1}{2i}{\tiny \barray{cc}0 &1 \\ 1 &0\earray}
$. Then parallel transport implies that~\footnote{The two signs correspond to the two half-planes:  the sign should change when we switch from one to the other. Remember that $\varphi = \pm(\vartheta-\pi)$, depending on the sign of the orbital angular momentum, $L$.},
\beq
\homega_{\sigma_1}(\varphi) = {Q}^p_{\pm}(\varphi) 
=\cos\varphi\frac{\sigma_1}{2i} \mp \sin \varphi\frac{\sigma_2}{2i}
=
\frac{1}{{2}i}\barray{cc}
0 &e^{\pm{}i\varphi}
\\
e^{\mp{i}\varphi}&0\earray\,.
\label{omegaproton}
\eeq  
The charge is the projection onto the direction field, 
 \eqref{omegacharge}. Spelling out for the proton trajectories \eqref{P+inout}-\eqref{P-inout} we get indeed,
for all $\varphi$
\beq
q_{proton} = -2\tr\left({Q}^p_{\pm}(\varphi)\right)^2
= 1\,,
\label{protoncharge1}
\eeq
as expected: $\sigma_1\in\su(2)$ generates an internal symmetry and thus the charge is conserved: the incoming proton remains a proton during the scattering.
For the S-matrix for example, the minus sign in \eqref{II-II} combines with the minus sign of the direction field and thus yield conserved charge equal to $1$, interpreted as the electric charge of the proton. 

For $\sigma_{2}$ and $\sigma_{3}$ we have instead, by \eqref{N+-inout}, 
\besub
\begin{align}
\homega_{\sigma_2}(\varphi) = &\; \frac{1}{2}
{\tiny \barray{cc}
0 &-e^{\pm{i}\varphi}
\\[6pt]
e^{\mp{i}\varphi}&0\earray}\,
\Rarrow
q_{neutron} = \;0\,,
\label{neutronfield}
\\[8pt]
\homega_{\sigma_3}(\varphi) = & \;\;\quad \frac{1}{2i}
\barray{cc}1 &0\\ 0 &-1\earray
\quad \Rarrow \;\;\, q_{vert} \;\;={Q^3(0)}\,.
\label{verttransl}
\end{align}
\label{zerocharges}
\esub
The 3rd charge here corresponds to $\sigma_3$   perpendicular to the plane of the motion. For our choice it vanishes.
These formul{\ae} are consistent with  \eqref{orthonorm} and confirm that when the 
flux is quantized in odd units, 
$
\Phi = -(1+2k)\pi({\sigma_3}/{i})
$ 
with $k$ an integer, cf. \eqref{oddflux}, then we have three conserved charges.  
However if ${\fa}$ is \emph{not an integer} \ie, for flux $\vPhi \neq N\pi\sigma_3$ the charge is \emph{not
conserved}. For ${\fa}=1/2$, for example, 
\beq
\left({Q}^p_{\half}\right)_{\pm}\!(\varphi) 
=
\frac{1}{{2}i}\barray{cc}
0 &e^{\pm{}i\varphi/2}
\\
e^{\mp{i}\varphi/2}&0\earray
\label{ahalfQ}
\eeq
and the charge is \emph{not} conserved,
\beq
\Rarrow
q_{\half}(\varphi)=
-2\tr\left(\homega_{\sigma_1}\cdot{Q}_{\half}\right)= 
\cos \frac{\varphi}{2}\neq \const\,.
\label{a=halfcharge}
\eeq  
Thus the \emph{incoming proton} looses its initial charge $q_{in}=1$ and goes out with zero charge, $q_{out}=0$, \ie, as a \emph{neutron},
\beq
Q_{in} = Q^{p} \to Q_{out} = Q^{n}\,,
\label{ahalfscatt}
\eeq
as argued by Wu and Yang \cite{WuYang75}.  

Intuitively, the ``electromagnetic direction field'' $\homega$ is in fact that of a proton for ${\fa}=1+2k$ (drawn in red and in blue in  FIG.\ref{clockwork}), and the isospin for ${\fa}=\half$ (drawn in magenta and cyan) turn at different angular velocity, therefore the projection of the latter onto the former one is not a constant, \eqref{a=halfcharge}. 
Similarly, an incoming neutron would come out as a proton,  as shown in FIG.\ref{clockwork}. 

\begin{figure}[h]\hskip-5mm
\includegraphics[scale=.53]{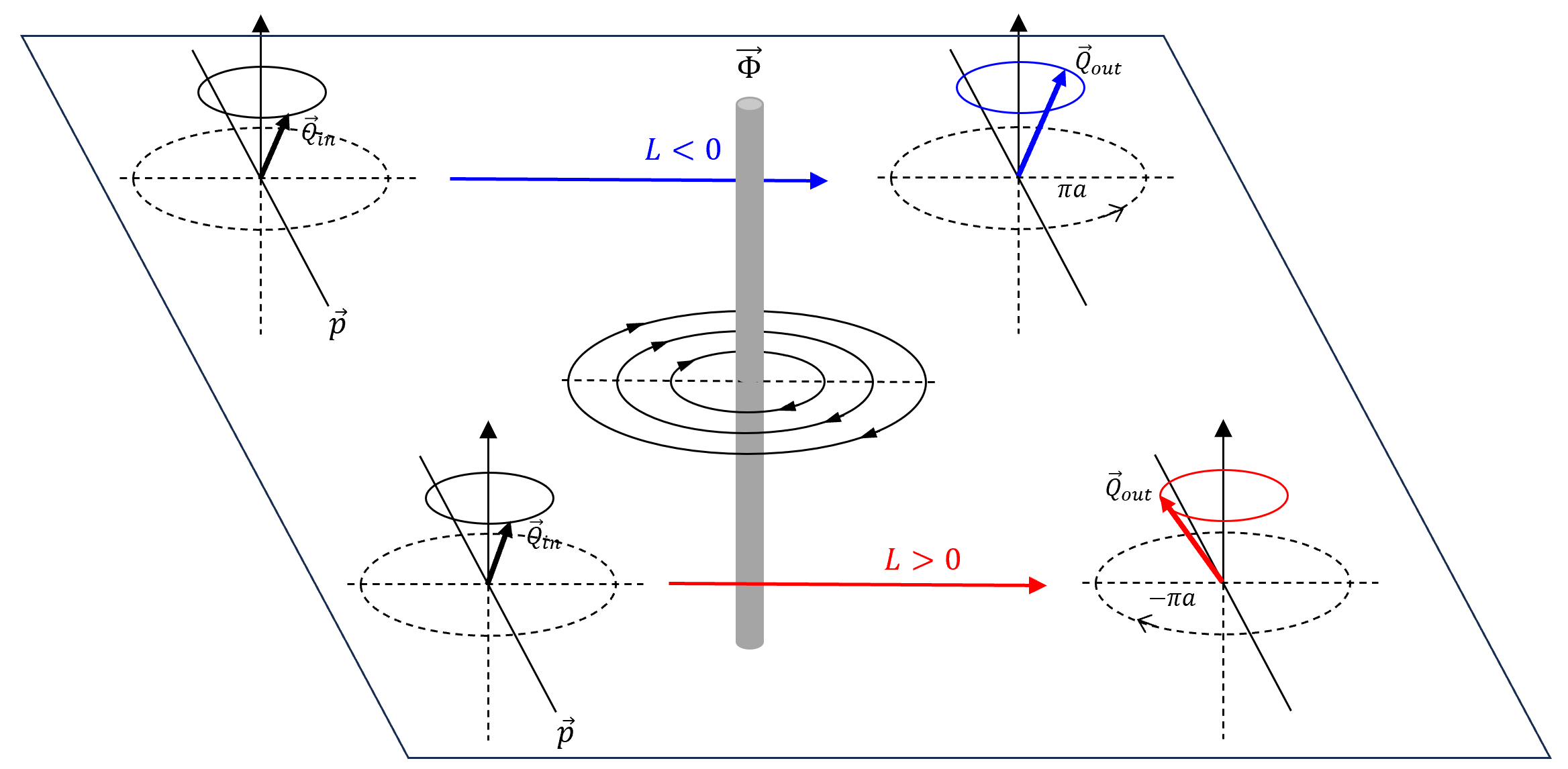}

\caption{\textit{\small The isospin vector $\vQ$ whose initial direction has an angle $\alpha$ with that of the non-Abelian flux $\vPhi$ rotates  along a horizontal circle on the cone of opening angle $\alpha$ around the flux' direction. The sense whose the rotation depends on the side the. The apex of the cone is carried along the free motion in space. The charge defined as the projection of $\vQ$  onto the comoving ``electric direction'' $\homega$ is conserved when the flux is quantized in odd units  \eqref{oddflux}. } 
\label{newproton}
}
\end{figure}

Generalizing our previous investigations, we now consider motion with the same external initial conditions
as before 
 but  allowing a non-zero vertical component for the isospin,
\beq
{Q}_{in}=\left(\sin\alpha \frac{\sigma_1}{2i}
+\cos\alpha \frac{\sigma_3}{2i}\right)\,,
\label{alphaincond}
\eeq  
where $0\leq\alpha\leq\pi$ is the angle of the initial isospin with the $\sigma_3$-oriented flux, shown in FIG.\ref{newproton}. 
 The spatial trajectory remains unchanged, whereas the isospin equation is solved by
\beq
{Q}_{\pm}(\varphi)= \frac{1}{2i}\left\{
\,{\sin\alpha}
\barray{cc} 0 &e^{\pm{i}\varphi{\fa}}
\\ 
e^{\mp{i}\varphi{\fa}}
 & 0\earray
+\cos\alpha \,\sigma_3\right\}\,.
\label{alphatraj}
\eeq
Then the projection onto the flux' direction is
\beq
\vec{{Q}}\cdot \vPhi= 
{Q}^3(0) = \cos\alpha = \const
\,,
\label{alphaangle}
\eeq
which implies that \emph{the isospin precesses} on a  circular section  at constant height of a cones with opening angle $\alpha$ w.r.t. the $\sigma_3$-oriented flux $\vPhi$, whereas theirs apexes  travel along the free spatial trajectories as for $\alpha = \pi/2$, see FIG.\ref{newproton}.
Maintaining our charge-based proton/neutron definition with direction field $\homega_{\sigma_1}$,
\beq
\tilde{q} = \sin\alpha
\label{alphacharge}
\eeq
which is conserved,  is 
 however neither a proton nor a neutron except for $\alpha=0$ (when it is a neutron) or $\alpha=\pi/2$ (when it is a proton)~\footnote{The situation is analogous to that observed for classical spin, which becomes a half-integer only under quantization \cite{SSD,HPAAix79}.}.
Similarly, the charge generated by $\homega_{\sigma_3}=\sigma_3$, 
\beq
-2\tr \left(\tilde{Q}_{\pm}(\varphi)\cdot\homega_{\sigma_1}\right)=\cos\alpha\,,
\eeq
although a constant of the motion, is $1$ only for $\alpha=0$ but is zero 
$\alpha=\pi/2$ \eqref{verttransl}.

However as emphasised by Yang and Mills, we are allowed to change our proton/neutron definition and call instead proton an incoming particle with 
$
\tilde{\Psi}(\varphi=0) = {Q}_{in}
$ 
in \eqref{alphaincond}
and promote \eqref{alphatraj} with ${\fa}=1$  to a ``moving reference frame''
\beq
\tilde{\Psi}(\varphi) = \frac{1}{2i}\left\{
\,{\sin\alpha}
\barray{cc} 0 &e^{\pm{i}\varphi}
\\ 
e^{\mp{i}\varphi}
 & 0\earray
+\cos\alpha \,\sigma_3\right\}\,,
\qquad
0\leq\varphi\leq\pi,
\label{newrefframe}
\eeq
Then  projecting onto the latter,
\beq
\tilde{q} =-2\tr\,(\tilde{\Psi}\cdot \tilde{Q}) = 
\frac{1}{2}\tr\,\left\{\sin^2\alpha \barray{cc}
e^{i\varphi({\fa}-1)} &0
\\
0 &e^{-i\varphi({\fa}-1)}
\earray +
\cos^2\alpha \barray{rrr}1 &\quad &0 \\ 0 &\quad &1 \earray\right\}\,,
\label{newcharge}
\eeq
would again be a satisfactory charge definition in that it would be (i) gauge-covariant and (ii) for ${\fa}=1$
equal to $1$ for all $\varphi$ by Pythagoras' theorem. 
The redefined  charge \eqref{newcharge} would again be conserved and remain a ``proton'' with respect  the new direction field $\tilde{\Psi}$.

The electric-type internal charges defined by projecting onto a covariantly constant moving reference frame, \eqref{omegacharge}, are conserved. 

Another curious feature of the inherent freedom advocated by Yang and Mills \cite{YangMills} to choose what we call  a proton \cite{WuYang75} is the following. Let us assume that  ${\fa}=1$; then we have 3 internal symmetries.  
Selecting again  the covariantly constant direction field $\homega_{\sigma_1}$ 
 \eqref{omegaproton} above provides us, in the twisted case, with 3 conserved charges as seen above.  
However, the components of a \emph{constant} isospin trajectory,
\beq
\vQ(\varphi) = \vQ_0 = \const\,,
\label{constQ}
\eeq 
which would provide us with trivially conserved charges in the even case $\fa = 2k$ will \emph{not}  be conserved in the twisted case $\fa=1$ in general. Using \eqref{NABAcharges} we find indeed,
\beq
\left\{\barraynb{llll}
Q^1(\varphi)= \frac{\sigma_1}{2i} &\qquad
&q^1=
&\cos\varphi\neq\const 
\\[5pt]
Q^2(\varphi) =\frac{\sigma_2}{2i} &\qquad
&q^2= &\sin \varphi \neq \const
\\[5pt]
Q^3(\varphi) =\frac{\sigma_3}{2i} &\qquad
&q^3 
= &0
=\const
\earraynb\right.
\label{2+1charges}
\eeq

\subsection{Spin from isospin}\label{Spinfromisospin}
  
To get an insight, we start with the trivial case ${\fa}=2k$ \ie, $\fF=\II$. The particle is free and 
the total angular momentum,
\beq
\vJ= \vL -\vQ\, \where \vL =\vr\times \vp\,,
\label{JpQ} 
\eeq
is conserved. The two terms here are in fact
\emph{separately conserved}~: $\vL$,
the orbital one is conserved, because the spatial motion is free; the internal one, $\vQ$, is conserved because in the diagonal gauge the commutator drops out from the isospin equation \eqref{isospineq} leaving us with $d\vQ/dt=0$.   For ${\fa}=2k$ the gauge potential can be gauged away (as noted above) and the equation is gauge-covariant.

This scenario realizes Souriau's ``d\'ecomposition barycentrique" \cite{SSD} chap. III. p.163 \footnote{Anticipated by Jacobi in his K\"onigsberg  lectures \cite{Jacobi}.}, which says that when the ``space of motions" (his replacement for the phase space) has at least 8 dimensions which is our case here because of the isospin, then, for an isolated system with pairwise internal interactions, the rotations of the center of mass and the rotations around the latter act separately and both contribute to the angular momentum.  
Thus the $\SO(3)$ symmetry is in fact enlarged to 
$ \SO(3)_{ext}\times\SO(3)_{int}\,.$ 

In the \NA \AB setup the flux is enclosed into a thin cylinder that arguably reduces the external symmetry to an axial $\SO(2)_{ext}$. However this mechanical constraint is unrelated to the internal structure (isospin) of our particles, leaving us with the enlarged rotational symmetry
\beq
\SO(2)_{ext}\times\SO(3)_{int}\,,
\label{doublerot}
\eeq
generated by the orbital momentum $\vL$ and the isospin $3$-vector $\vQ$, respectively, as in \eqref{JpQ}.

Souriau illustrates his ``th\'eor\`eme de d\'ecomposition barycentrique'' by a non-relativistic particle with spin \cite{SSD}. However, expressed in Souriau's language, the internal symplectic form for the \emph{isospin} is in fact \emph{identical} to that for his classical non-relativistic spin \cite{SSD}, namely the surface form of the two-sphere of radius $|Q_0|$, $\cO\sim \IS^2 \subset \su(2)$.
The (separately) conserved quantities are the  orbital angular momentum $\vL$ and the 3 internal charges which are in fact the 3 conserved components $Q^i$ of the isospin  in eqn. \eqref{NABAcharges}. 

What happens in the non-trivial case ?
Our general theory implies that when ${\fa}$ is not an integer
then no covariantly constant ``direction fields" 
and thus no internal symmetries do exist,
 implying that the \emph{total angular momentum \eqref{JpQ} 
 can not be conserved, leaving us with the mere orbital expression.} 
 
 We are still left with the other exceptional case, ${\fa}=1+2k$. However recall that the \emph{orbital} angular momentum, $\vL$, \emph{is} conserved for \emph{all} ${\fa}\neq 2k$ as we have seen in Sec.\ref{CCharge}.
 But this implies that the total angular momentum, \eqref{JpQ} can not be conserved whenever the isospin is not a constant of the motion (except for a particular  initial condition). We  conclude that the internal structure \emph{breaks}  \emph{the full rotational symmetry \eqref{JpQ} to mere orbital one:}
 for ${\fa}=1+2k$, $\sigma_3$ generates, as before, internal rotations around the flux' direction 
which yields the  internal angular momentum,
\beq
L_{int} = Q^3 \,\frac{\sigma_3}{2i}\,
\label{IntAngMom} 
\eeq
which is conserved since $\homega_{\sigma_3}=\sigma_3$ and $Q^3 = Q^3(0)$).
Thus the non-trivial flux breaks  the symmetry
\eqref{doublerot} down to a double \emph{axial} one, 
\beq
 \SO(2)_{ext}\times\SO(2)_{int}\,,
\label{double2rot}  %
\eeq
generated by \emph{independent external and internal rotations} around the (respective) 3rd directions.
For $\alpha=0$ the cone degenerates to a  vertical vector of unit length which is conserved; for $\alpha=\pi/2$ we recover our previous result in FIGs.\ref{clockwork}-\ref{a=1PN}-\ref{protoncyclo}.

The \emph{total angular momentum} \eqref{JpQ} is conserved in the vacuum case $a=2k$ ($\fF=\II$) but broken to \eqref{double2rot}  
for ${\fa}=2k+1$ ($\fF=-\II$),
leaving us with the 3rd component only,
\beq
J_3 = L_3 - Q^3\,.
\label{axangmom}
\eeq
The spin from isospin term (although only along the 3rd direction) but the two other components in \eqref{JpQ} are not more conserved --- the isospin precesses.
\goodbreak

\section{Conserved quantities by van Holten's algorithm}\label{vHoltenSec}

 To deepen our insight, we revisit the internal symmetry question from a rather different point of view, namely by using van Holten's approach \cite{vHolten,HP-NGOME}. His algorithm
 amounts to developping the conserved quantity we are looking for into powers $\Pi_{i} = \dot{x}_{i} $ (we took $m=1$),
\begin{equation}
I=C+C_{i}\Pi _{i}+\frac{1}{2}C_{ij}\Pi_{i}\Pi_{j}+...
\label{Idevelop}
\end{equation}%
whose conservation, $\{I,H\}=0$ requires 
\begin{eqnarray}
&& order ~ 1 : \qquad D_{i}C= Q^aF_{ij}^{a}C_{j} 
\nonumber 
\\
&& order ~ 2 : \qquad D_i C_j + D_j C_i  = Q^{a}\left(F_{ik}^{a}C_{kj}+F_{jk}^{a}C_{ki}\right)\,, \nonumber
\\
&& ...
\label{vHisoconstr}
\end{eqnarray}
which stops when we have a Killing tensor.
Applied to the \NABA\!,  $F_{ij}^{a}=0$, this simplifies to
\begin{eqnarray}
&& order ~ 1 : \qquad D_i C  = 0, \nonumber \\
&& order ~2 : \qquad D_i C_j + D_j C_i =0\,, 
 \nonumber \\
&& ...
\end{eqnarray}
For simplicity we restrict our attention at planar real-space motion,
  \ie, assume that the initial conditions imply that the real-space motion is in the $x-y$ plane with the $z$-direction identified with 
   $\sigma_3$ in isospace, $\vn \sim \sigma_3$.
Planar rotations, for example,
are generated by  the Killing vector  
$ 
\vn\times \vx\,.
$ Then
the algorithm yields  $C = 0,\, C_{i}=-\varepsilon_{ij}x^{j}$ 
 \cite{vHolten,HP-NGOME} and we recover the 3rd component of the \emph{conserved orbital angular momentum} \eqref{Lorbital}, 
$ L \equiv L_3=\varepsilon_{ij}x^{i}\dot{x}^j$\,,
as expected. 

 Now we rederive our previous results about internal charges by the van Holten method. Searching for the first-order in $\vQ$ coefficient in the form 
\begin{equation}
C\left(\vec{x},Q\right) =C^{a}\left( \vec{x}\right) Q^{a}\, 
\label{ClinearQ}
\end{equation}%
and using the parallel transport eqn. $D_{i}Q^{a}=0$, %
we get
\begin{equation}
D_{i}C^{a}=\partial _{i}C^{a}-\varepsilon^{abc}A_{i}^{b}C^{c}=0\,.
\label{PM17}
\end{equation}%
Inserting here the \NA \AB gauge potential
$
A_{i}^{a}=
-\frac{B^{a}}{%
r^{2}}\varepsilon_{ij}x^{j}
$
with $(B^a) =(0,0,{\fa})$, these equations become
\begin{equation}
\left\{
\barraynb{lllll}
\partial_{x}C^{1}-aC^{2}\frac{y}{r^{2}}=0,&\qquad
&\partial _{x}C^{2}+aC^{1}\frac{y}{r^{2}}=0,&\qquad
&\partial _{x}C^{3}=0,
\\[6pt]
\partial_{y}C^{1}+aC^{2}\frac{x}{r^{2}}=0,&\qquad
&\partial_{y}C^{2}-aC^{1}\frac{x}{r^{2}}=0,&\qquad
&\partial_{y}C^{3}=0\,,
\earraynb%
\right.
\end{equation}
which admits the nontrivial solution 
\begin{equation}
C^{1}(x,y)=\alpha\sin\mu,~C^{2}(x,y)=\alpha\cos\mu,
\where \mu = \mu(x,y) = {\fa}\arctan\frac{x}{y}+\phi_{0}\,,
\label{solu1}
\end{equation}%
where $\phi_{0}$ is an arbitrary constant. This should be
completed with $C^{3}=\beta=\const$ which follows from having  directed the vector potential into the 3rd direction.
\goodbreak

For $\alpha =0$, the isospin has only a $\sigma_{3}$ component and 
\eqref{IntAngMom} is recovered. We put hence $\alpha=1,\,\beta =0$ and focus our attention at the two remaining components.
Then the algorithm yields, 
\beq
I = \sin\mu \,Q^{1} + \cos\mu \,Q^{2}  \,,
\label{PM(26)}
\eeq%
This eqn.
 is plainly  valid for $y\neq 0$,  \ie, in the upper and lower half-planes separated by the $x$-axis because of 
$y$ in the denominator for $\mu$. 
However letting $y \to 0$ from the two sides yields, for fixed $x>0$, $x/y \to \pm \infty$, depending on the sign of $y$.  Then 
 the asymptotic behavior of the $\arctan$ implies, 
\beq
\mu \to 
\left\{
\barraynb{lrlll} 
\mu_+ =
&\;{{\fa}\pi}/{2}+\phi_0  &\when  &y>0
\\[3pt]
\mu_- = &- {{\fa}\pi}/{2}+\phi_0  &\when   &y < 0
\earraynb\right.
\Rarrow \mu_- =\mu_+- {a\pi}\,.
\label{y+=}
\eeq
Thus the trigonometric functions do \emph{not} match in general, $I_+\neq I_-$~:  
there is \emph{no well-defined conserved quantity on the entire plane}, consistently with the  breaking of the internal symmetries found earlier.

However remember that there might be a \emph{transition function} (in fact a constant element) between the two half-planes which should belong to the center of the gauge group \cite{HR-PRD,HR-color},
\beq
I_{+}=h\,I_{-}\,,
\where
h \in Z_{G}= \{\II, - \II\}\,.
\label{I+I-}
\eeq

For $h = \II$ this requires $\sin\frac{{\fa}\pi}{2}=0$ so that ${\fa}$ is an even integer,
$
{\fa} = 2k.
$
For $h = -\II$ we get instead $\cos\frac{{\fa}\pi}{2}=0$ implying that $a$ is an odd integer,
$
{\fa} = 1+2k\,.
$
In both cases we get from 
 by using \eqref{ClinearQ} with \eqref{solu1} \eqref{PM(26)},
which is a well-defined conserved quantity for both of  exceptional values of ${\fa}$.

Conversely, remembering that the length of $\vQ$ is a constant, $|\vQ|^2 = I^2$,
allows us to recover the trajectory using the conserved quantity $I$ as \footnote{More generally, 
$|\vQ|^2 = I^2+|Q^3|^2$.}
\beq
Q^1(x,y) = I \sin \mu (x,y) 
\aand
Q^2(x,y) = I \cos \mu(x,y) \, 
\label{PMQ}
\eeq
consistently with \eqref{P+inout}-\eqref{P-inout} and \eqref{N+-inout}.
Inserting \eqref{PMQ} into \eqref{PM(26)} identity as it should. 
The conserved quantity  \eqref{PM(26)} is indeed the conserved charge $q=Q^{a}\Psi^{a}$ in \eqref{qdef} or more generally, in \eqref{NABAcharges} with  \eqref{PMQ} for $\homega_{\sigma_1}$ chosen as in \eqref{omegaproton}. 

In particular for ${\fa}=2k$ (even)  $\mu \to {\pm}k\frac{\pi}{2} + \phi_{0}=\const$, implying that $\vQ$ is a constant,
\beq
Q^1 = (-1)^k\,I\,\sin\phi_0 = \const
\aand
Q^2 = (-1)^k\,I\,\cos\phi_0 = \const\,.
\label{k0Q}
\eeq

For ${\fa}=1+2k$ (odd) we have  $\mu = \mu(x,y)\neq\const$, \eqref{PM(26)}, however the limits from the two sides, 
$\mu_{\pm} \to \pm \frac{\pi}{2}(1+2k) + \phi_0$, do match when $y\to0$, \eqref{y+=}.

These results confirm those found in sec.\ref{ScattSec}. as we show for ${\fa}=1+2k$ with
 $y>0$. For the ``in" (resp. ``out'') states,  
choosing $\phi_0 =(1+k)\pi$ \eqref{PMQ}  implies, 
\beq
Q_{in}^1 = 1,\;\; Q_{in}^2 = 0
\quad\text{\small resp.}\quad
Q_{out}^1 = -1,\;\; Q_{out}^2 = 0\,,
\label{Qinout1}
\eeq
consistently with \eqref{outa=1}. The other values shown in FIG.\ref{a=1PN} are recovered also by appropriate choices of $\phi_0$.
\goodbreak

\medskip
Returning to the angular momentum, we get, consistently with our previous results,

\begin{thm}
When the Wu-Yang factor does not belong to the center, $\fF \neq\pm\II$, then
 only the 3rd component of the orbital angular momentum, $L=L_3$,  is conserved. However when the flux is {quantized in either of the two series  in \eqref{NAfluxquant}}, 
then $L_3$ and $Q^3$ in \eqref{Lorbital} and \eqref{IntAngMom}, are both separately conserved and the symmetry is doubled, \eqref{double2rot}, consistently with Souriau's {\sl d\'ecomposition barycentrique} \cite{SSD}.
\label{vHsymmthm}
\end{thm}
\goodbreak

\section{Conclusion and outlook}\label{Concl}

This project, suggested by T. T. Wu to PAH over 40 years ago \cite{NABA82,NABAPRD,HP-EPL,HP-Kollar1,HP-Kollar2,Sundrum}{} \footnote{
H. Bacry, musing over a piece of paper in Marseille in the eighties, more than once whispered : \emph{``On \'ecrit le m\^eme article durant toute sa vie''} (One writes the same paper during all his life).}, 
  studies the \NABA proposed by Wu and Yang in their celebrated paper on the non-integrable phase factor \cite{WuYang75}.
While the original proposal is  field-theoretical/quantum mechanical, here we argue that a ``shadow'' of it, namely \emph{isospin precession}\footnote{Isospin precession was found also for diatomic molecules \cite{MSW,Jackiw86}; see the recent study \cite{KernerFest}.}, can be recovered also at the the pseudo-classical level, with the isospin represented by a vector $\vQ$ of the Lie algebra of the gauge group \footnote{Subtleties of the concept of classical isospin were discussed in \cite{Arodz82,Arodz82b,Arodz83}.}.
 This is in sharp contrast with the Abelian (electromagnetic) case~: \AB effect is purely quantum mechanical \cite{AharonovBohm}. 
 
Comparing the Abelian and the \NA case, we argue that 
``the electric charge is a \emph{constant scalar} (in fact the same constant for all motions), but the non-Abelian  isospin \emph{vector} has a proper dynamics, -- even if it is rather poor and, being parallel transported, is determined by the space-time motion.
Thus the difference comes from the (partial) observability of the phase. 

Physical interpretation in terms of protons and neutrons 
can be obtained by following the charge-based arguments of Yang and Mills \cite{YangMills}. The charge  is the projection of the isospin vector onto a chosen direction field $\hPsi(x^\mu)$, \eqref{omegacharge}. Then a proton is a particle with charge $1$ and a {neutron} has zero charge.
However, the charge is conserved only when $\hPsi(x^\mu)$ is covariantly constant \eqref{ocovconst}, 
$D_\mu\hPsi=0$,
\ie, when $\hPsi$ corresponds to an internal symmetry \cite{HR-PRD,HR-color}.

For non-Abelian monopoles this happens outside their core \cite{tHooft,Polyakov,GoddardOlive}, allowing to reduce the dynamics of a test particle to an Abelian problem \cite{Feher:1984ik,Feher:1984xc}.

In the \NA \AB case, though, a covarantly constant $\hPsi$ exists only  when the \NA \emph{flux is quantized} in either of the two distinct series \eqref{NAfluxquant}, 
  then we get 3 conserved internal charges \cite{HR-PRD,HR-color}; for other flux values we do not get any.    
The flux quantization we find here is reminiscent of the one in superconducting rings \cite{DeaverFairbank,ByersYang,Onsager}.

This surpising behaviour hints at the \emph{global structure} of the gauge group: it comes from the center of $\SU(2)$ being composed of $\II$ and $-\II$; for $\SO(3) = \SU(2)/\IZ_2$ we have only $\II$. Thus answering the question~: --- ``Is the non-Abelian flux quantized ?  And if it is, then in which units~?'' would depend on what the ``true'' gauge group is.

The most dramatic prediction of Wu and Yang \cite{WuYang75} is that an incoming proton might be turned into an outgoing neutron under specific circumstances. 
 We confirm their statament when the  flux is not quantized~: for half-integer flux ${\fa}=\half$, for example, then $D_{\mu}\hPsi\neq0$ and an \emph{incoming proton could indeed turn into an outgoing neutron}.

However the initial choice can be changed yielding different casts,
 illustrating the arbitrariness of identifying protons and neutrons, as emphasised by Yang and Mills \cite{YangMills}.

Such an eventuality raises the fundamental question :  Is such a transition indeed possible or not ? What would happen with the electric charge ? What could be the physical mechanism for it. We confess not to be able to answer. It might well happen that the test particle framework we follow in this paper (or its quantum mechanical antecedents \cite{NABA82,HP-Kollar1,HP-Kollar2,HP-EPL}) are just not sufficient to  answer. Another idea could be to enlarge the commutative part of the gauge group from $\SU(2)$  to Weinberg-Salam $\UU(2)$ (locally $(\UU(2)\times \UU(1)$).

The direction field $\hPsi(x^\mu)$ behaves, mathematically,  as a  \emph{non-dynamical Higgs field} in the adjoint representation -- which
however  plays only a passive role, as a sort of ``measuring apparatus''. 
 For a Wu-Yang monopole \cite{WuYang69}, the archi-type of non-Abelian monopoles, for example, it corresponds to their radial ``hedgehog'' direction \eqref{hedgehogHiggs}. 
For a 't Hooft-Polyakov monopole instead, the direction field is promoted to a dynamical field which allows for finite-energy soliton solutions of the coupled Yang-Mills-Higgs equations \cite{GoddardOlive,tHooft,Polyakov}. The direction field chosen tacitly  is the normalized physical Higgs field, $\homega=\Psi/|\Psi|$.
 Outside the monopole core the $\SU(2)$ gauge symmetry is spontaneously broken to a residual $\UU(1)$ implying that the Higgs field is covariantly constant there.

The \NA flux line we studied in this paper corresponds, intuitively, to a pure YM vortex whose ``outside region'' is the entire plane with the origin removed.
Form the point of view of a field-theoretical YMH Lagrangian, the kinetic  term $D_{\mu}\Psi D^{\mu}\Psi$ is switched off, reducing $\Psi$ to a non-dynamical quantity.

The contrast between pure YM and YMH becomes even more salient in Chern-Simons theory \cite{JackiwWeinberg,JackiwPiCS}, whose non-Abelian version \cite{DJPT,D-LNP} is strongly reminiscent of  \NA \AB --- except for that the Higgs field is, once again, dynamical. However the  analytic solutions found in \cite{DJPT,D-LNP} are as remote as possible from the \NA \AB in that they are \emph{self-dual},
$ 
\half\epsilon_{ijk}F^{jk}= D_i\Psi\,.
$ 
instead of our cast with $F^{jk} \propto \delta(x)$.
 
 The angular momentum in the \NABA exhibits the ``spin from isospin'' phenomenon \cite{JackiwRebbi76,Hasi,JackiwManton,DH82,Chen:2008ag}.
The orbital angular momentum is conserved  for all fluxes. However for the two particular values of the flux in 
\eqref{NAfluxquant}, we have additional terms generated by  internal symmetry which is \emph{independent} of space-time rotations. The total angular momentum, 
$J$ in \eqref{JpQ} splits into two separately conserved pieces, one external (orbital) and one internal (isospin),
realizing
Souriau's  doubled rotational symmetry idea \cite{SSD}.
When the flux quantized in odd units, eqn \eqref{oddflux}, 
the remnant of the internal symmetry is detected by the precessions of the isospin on horizontal circles, shown in FIG.\ref{newproton}.

Further insight can be gained by looking at the problem from a different point of view, advocated by van Holten \cite{vHolten,HP-NGOME}, see sec.\ref{vHoltenSec}. 

The flux quantization we found above has some analogy with  
the Dirac quantization of a monopole: choosing the Dirac string along (say)  the negative $z$ axis,  in its neighborhood 
the vector potential will behave approximately as an \AB
solenoid, which remains unobservable when the flux is quantized \cite{GoddardOlive,FrenkelHrasko,HraskoBalog}.

Thus the \NA \AB setup is indeed the planar analog of the \NA monopole  \cite{GoddardOlive,GNO,HRCMP1}. 
Our investigations confirm the \emph{credo} of  Wu and Yang \cite{WuYang75}~:

\begin{quote}\textit{\narrower 
\dots what constitutes an intrinsic and complete description of electromagnetism? [\dots] An examination of the Bohm-Aharonov experiment indicates that in fact only the phase factor \eqref{ABPF}  and not the phase is physically meaningful. 
}
\end{quote}

Quantum aspects were studied in \cite{NABA82,HP-Kollar1,HP-Kollar2,HR-PRD,HR-color,NABAPRD,HP-EPL,Sundrum}. See \cite{KernerFest,Bright15,Hosseini16,Chatterjee15,Cserti} for a (very incomplete) selection. 
It is remarkable that  part of  the \NABA can be captured by the (pseudo-)classical isospin of Kerner and Wong \cite{Kerner68,Wong70}, confirming that the isospin is neither a truly classical nor a truly quantum concept: it is at half ways between the two, justifying our terminology.  

Yang et al \cite{YYang,MITNews} attribute the \NA \AB effect to the breaking of time-reversal symmetry. However parallel transport implies that real time is replaced, in our framework, by the angle $\varphi$. Thus the effect comes from ``navigating" on either sides of the flux. 
However we found that when the flux is quantized, \eqref{NAfluxquant} then the 
time reversal symmetry disappears: side-dependent phase shifts yield identical contributions on both sides,
as highlighted by doubling the rotational symmetry, as in \eqref{JpQ}-\eqref{double2rot}. 
 
\smallskip
Intensive ongoing work on effective/artifical gauge fields  
\cite{MSW,Jackiw86,KernerFest,Bright15,Hosseini16,Chatterjee15,Cserti,Osterloh,Goldman,Jacob,Dalibard,YChen,YYang,Fruchart,YBiao,YangYang,NABA-PR} provide promising testing ground of
the conceptual suggestions of  Wu and Yang, 
allowing us to hope that the \NABA may, on one day, not be considered anymore as \dots ``dry water'' (using a celebrated expression put forward by J. von Neumann). 

\vskip-3mm
\begin{acknowledgments} 

\smallskip
This paper is dedicated to the memory of ``Tai'' (Tai-Tsun Wu : 1933-2024)).
Correspondance with Janos Balog, Gary Gibbons, Csaba S\"uk\"osd and A. Zeilinger is gratefully acknowledged.
PMZ was partially supported by the National Natural Science Foundation of China (Grant No. 11975320). 
\end{acknowledgments}
\goodbreak



\begin{thebibliography}{99}

\bibitem{AharonovBohm}
Y.~Aharonov and D.~Bohm,
``Significance of electromagnetic potentials in the quantum theory,''
Phys. Rev. \textbf{115} (1959), 485-491
doi:10.1103/PhysRev.115.485

\bibitem{Chambers}
R.~G.~Chambers,
``Shift of an Electron Interference Pattern by Enclosed Magnetic Flux,''
Phys. Rev. Lett. \textbf{5} (1960) no.1, 3-5
doi:10.1103/physrevlett.5.3;

\bibitem{Mollen}
G. M\"ollenstedt and W. Bayh, 
``Messung der kontinuierlichen Phasenschiebung von Elektronenwellen im kraftfeldfreien Raum durch das magnetische vektorpotential einer Luftspule,''
Naturwiss. \textbf{49}, 81 (1962)
https://doi.org/10.1007/BF00622023.

\bibitem{Osakabe}
N.~Osakabe, T.~Matsuda, T.~Kawasaki, J.~Endo, A.~Tonomura, S.~Yano and H.~Yamada,
``Experimental confirmation of Aharonov-Bohm effect using a toroidal magnetic field confined by a superconductor,''
Phys. Rev. A \textbf{34} (1986), 815-822
doi:10.1103/PhysRevA.34.815

\bibitem{Peshkin89}
M.~Peshkin,
``The Aharonov-Bohm Effect. Part 1: Theory,''
Lect. Notes Phys. \textbf{340} (1989), 1-34
doi:10.1007/BFb0032077

\bibitem{Tonomura87}
A. Tonomura, 
``Applications of electron holography''
Rev. Mod. Phys. 59, 639-669 (1987).
DOI:https://doi.org/10.1103/RevModPhys.59.639

\bibitem{Tonomura89}
A.~Tonomura,
``The Aharonov-Bohm Effect. Part 2: Experiment,''
Lect. Notes Phys. \textbf{340} (1989), 35-152
doi:10.1007/BFb0032078

\bibitem{TonomuraToday}
H.~Batelaan and A.~Tonomura,
``The Aharonov-Bohm effects: Variations on a subtle theme,''
Phys. Today \textbf{62N9} (2009), 38-43
doi:10.1063/1.3226854

\bibitem{WuYang75}
T.~T.~Wu and C.~N.~Yang,
``Concept of Nonintegrable Phase Factors and Global Formulation of Gauge Fields,''
Phys. Rev. D \textbf{12} (1975), 3845-3857
doi:10.1103/PhysRevD.12.3845

\bibitem{Schulman71}
L. S. Schulman,
``Approximate topologies''
J. Math. Phys. \textbf{12}, 304 (1971)
https://doi.org/10.1063/1.1665592

\bibitem{Schulman81}
L. S. Schulman,
{\sl Techniques and applications of path integration}, J. Wiley, N.Y. (1981);

\bibitem{Laidlaw}
M.G.G. Laidlaw and C. Morette-DeWitt,
Phys. Rev. \textbf{D3}, 1375 (1971);

\bibitem{Dowker72}
J.S. Dowker, 
J.Phys. \textbf{A5}, 936 (1972);

\bibitem{Dowker79}
J.S. Dowker,
{\sl ``Selected topics in topology and quantum field theory,''}, Austin Lectures (1979)

\bibitem{Horvathy88}
P.~A.~Horvathy, G.~Morandi and E.~C.~G.~Sudarshan,
``Inequivalent quantizations in multiply connected spaces,''
Nuovo Cim. D \textbf{11} (1989), 201-228
doi:10.1007/BF02450240

\bibitem{Horvathy79}
P.~A.~Horvathy,
``Quantization in Multiply Connected Spaces,''
Phys. Lett. A \textbf{76} (1980), 11
doi:10.1016/0375-9601(80)90133-4

\bibitem{Horvathy80}
P.~A.~Horvathy,
``Prequantization From Path Integral Viewpoint,''
Lect. Notes Math. \textbf{905} (1982), 197-206
CPT-80-P-1230. Available as [arXiv:2402.17629 [math-ph]].

\bibitem{Kobayashi}
S. Kobayashi and K. Nomizu, {\sl Foundations of Differential Geometry} Vols. I and II, (Interscience, New York, 1963 and 1969)

\bibitem{Husemoller}
D. Husemoller, {\sl Fibre Bundles}. 3rd edition Graduate Texts in Mathematics. https://doi.org/10.1007/978-1-4757-2261-1
Springer NY (2013).

\bibitem{YangMills}
C.~N.~Yang and R.~L.~Mills,
``Conservation of Isotopic Spin and Isotopic Gauge Invariance,''
Phys. Rev. \textbf{96} (1954), 191-195
doi:10.1103/PhysRev.96.191

\bibitem{Zeilinger}
A.~Zeilinger, M.A Horne and C.G. Shull, 
``Search for unorthodox phenomena by neutron interference experiments,''
in {\sl Proc. Int. Symp. on the foundations of quantum mechanics}, Tokyo 1983, ed. S. Kamefuchi, (Physical Society of Japan), Tokyo 1984, pp.289-293.

\bibitem{NABA82}
P.~A.~Horvathy,
``The non-Abelian Aharonov-Bohm effect,''
BI-TP-82/14. Inspire code {Horvathy:1982fx}.
Available as arXiv: 2312.16133 [hep-th].

\bibitem{Schwarz82}
A.~S.~Schwarz,
``FIELD THEORIES WITH NO LOCAL CONSERVATION OF THE ELECTRIC CHARGE,''
Nucl. Phys. B \textbf{208} (1982), 141-158
doi:10.1016/0550-3213(82)90190-0

\bibitem{NelsonMano83}
P.~C.~Nelson and A.~Manohar,
``Global Color Is Not Always Defined,''
Phys. Rev. Lett. \textbf{50} (1983), 943
doi:10.1103/PhysRevLett.50.943

\bibitem{MarmoBala82}
A.~P.~Balachandran, G.~Marmo, N.~Mukunda, J.~S.~Nilsson, E.~C.~G.~Sudarshan and F.~Zaccaria,
``Monopole Topology and the Problem of Color,''
Phys. Rev. Lett. \textbf{50} (1983), 1553
doi:10.1103/PhysRevLett.50.1553

\bibitem{NelsonColeman84}
P.~C.~Nelson and S.~R.~Coleman,
``What Becomes of Global Color,''
Nucl. Phys. B \textbf{237} (1984), 1-31
doi:10.1016/0550-3213(84)90013-0

\bibitem{HR-PRD}
P.~A.~Horvathy and J.~H.~Rawnsley,
``Internal Symmetries of Nonabelian Gauge Field Configurations,''
Phys. Rev. D \textbf{32} (1985), 968
doi:10.1103/PhysRevD.32.968 ;

\bibitem{HR-color}
P.~A.~Horvathy and J.~H.~Rawnsley,
``The Problem of `Global Color' in Gauge Theories,''
J. Math. Phys. \textbf{27} (1986), 982
doi:10.1063/1.527119

\bibitem{NABAPRD}
P.~A.~Horvathy,
``The Nonabelian {Aharonov-Bohm} Effect,''
Phys. Rev. D \textbf{33} (1986), 407-414
doi:10.1103/PhysRevD.33.407

\bibitem{WILCZEKCOLEMAN}
M.~G.~Alford, K.~Benson, S.~R.~Coleman, J.~March-Russell and F.~Wilczek,
``The Interactions and Excitations of Nonabelian Vortices,''
Phys. Rev. Lett. \textbf{64} (1990), 1632
[erratum: Phys. Rev. Lett. \textbf{65} (1990), 668]
doi:10.1103/PhysRevLett.64.1632

\bibitem{Wilczek89}
F.~Wilczek and Y.~S.~Wu,
``SPACE-TIME APPROACH TO HOLONOMY SCATTERING,''
Phys. Rev. Lett. \textbf{65} (1990), 13-16
doi:10.1103/PhysRevLett.65.13

\bibitem{Alford90}
M.~G.~Alford, J.~March-Russell and F.~Wilczek,
``Discrete Quantum Hair on Black Holes and the Nonabelian Aharonov-Bohm Effect,''
Nucl. Phys. B \textbf{337} (1990), 695-708
doi:10.1016/0550-3213(90)90512-C

\bibitem{Preskill90}
J.~Preskill and L.~M.~Krauss,
``Local Discrete Symmetry and Quantum Mechanical Hair,''
Nucl. Phys. B \textbf{341} (1990), 50-100
doi:10.1016/0550-3213(90)90262-C

\bibitem{Brandenberger93}
R.~H.~Brandenberger,
``Topological defects and structure formation,''
Int. J. Mod. Phys. A \textbf{9} (1994), 2117-2190
doi:10.1142/S0217751X9400090X
[arXiv:astro-ph/9310041 [astro-ph]].

\bibitem{MSW}
J. Moody, A. Shapere, and F. Wilczek, ``Realization of magnetic monopole gauge fields : diatoms and spin precession,'' Phys. Rev. Lett. \textbf{56}, 893 (1986)

\bibitem{Jackiw86}
R. Jackiw, ``Angular momentum for diatoms described by gauge fields'' Phys. Rev. Lett. \textbf{56}, 2779 (1986).

\bibitem{KernerFest}
P.~A.~Horvathy and P.~M.~Zhang,
``Kerner equation for motion in a non-Abelian gauge field,''
{\sl The Languages of Physics.} A Themed Issue in Honor of Professor Richard Kerner on the Occasion of His 80th Birthday.
Universe \textbf{9} (2023) no.12, 519
doi:10.3390/universe9120519
[arXiv:2310.19715 [math-ph]].

\bibitem{Bright15}
M.~Bright and D.~Singleton,
``Time-dependent non-Abelian Aharonov-Bohm effect,''
Phys. Rev. D \textbf{91} (2015) no.8, 085010
doi:10.1103/PhysRevD.91.085010
[arXiv:1501.03858 [hep-th]],

\bibitem{Hosseini16}
S.~A.~Hosseinin Mansoori and B.~Mirza,
``Non-Abelian Aharonov\textendash{}Bohm effect with the time-dependent gauge fields,''
Phys. Lett. B \textbf{755} (2016), 88-91
doi:10.1016/j.physletb.2016.02.004
[arXiv:1601.08164 [quant-ph]].

\bibitem{Chatterjee15}
C.~Chatterjee and M.~Nitta,
``Aharonov-Bohm Phase in High Density Quark Matter,''
Phys. Rev. D \textbf{93} (2016) no.6, 065050
doi:10.1103/PhysRevD.93.065050
[arXiv:1512.06603 [hep-ph]].

\bibitem{Cserti}
R.~N\'emeth and J.~Cserti,
``Differential scattering cross section~of the non-Abelian Aharonov-Bohm effect in multiband systems,''
Phys. Rev. B \textbf{108} (2023) no.15, 155402
doi:10.1103/PhysRevB.108.155402
[arXiv:2306.13448 [quant-ph]].

\bibitem{Osterloh}
K. Osterloh, M. Baig, L. Santos, P. Zoller, and M. Lewenstein,
``Cold Atoms in Non-Abelian Gauge Potentials: From the Hofstadter `Moth' to Lattice Gauge Theory,''
Phys. Rev. Lett. \textbf{95}, 010403 (2005).

\bibitem{Goldman}
N.~Goldman, G.~Juzeli\={u}nas, P.~\"Ohberg and I.~B.~Spielman,
``Light-induced gauge fields for ultracold atoms,''
Rept. Prog. Phys. \textbf{77} (2014) no.12, 126401
doi:10.1088/0034-4885/77/12/126401
[arXiv:1308.6533 [cond-mat.quant-gas]].

\bibitem{Jacob}
A. Jacob,  P.~\"Ohberg, G. Juzeli\={u}nas, and L. Santos,
``Cold atom dynamics in non-Abelian gauge fields,'' 
Appl. Phys. B 89, 439 (2007).

\bibitem{Dalibard}
J. Dalibard, F. Gerbier, G. Juzeli\={u}nas, P.~\"Ohberg, 
``Colloquium: Artificial gauge potentials for neutral atoms,''
Rev. Mod. Phys. \textbf{83}, 1523-1543 (2011).

\bibitem{YChen}
Y.~Chen, R.~Y.~Zhang, Z.~Xiong, Z.~H.~Hang, J.~Li, J.~Q.~Shen and C.~T.~Chan,
``Non-Abelian gauge field optics,''
Nature Commun. \textbf{10} (2019) no.1, 3125
doi:10.1038/s41467-019-10974-8
[arXiv:1802.09866 [physics.optics]].

\bibitem{YYang}
Y.~Yang, C.~Peng, D.~Zhu, H.~Buljan, J.~D.~Joannopoulos, B.~Zhen and M.~Solja\v{c}i\'c,
``Synthesis and Observation of Non-Abelian Gauge Fields in Real Space,''
Science \textbf{365} (2019), 1021
doi:10.1126/science.aay3183
[arXiv:1906.03369 [physics.optics]].

\bibitem{Fruchart}
Michel Fruchart, Yujie Zhou and Vincenzo Vitelli,
``Dualities and non-Abelian mechanics,''
Nature 577, 636?640 (2020)

\bibitem{YBiao}
J. Wu, Z. Wang, Y. Biao, F. Fei, S. Zhang, Z. Yin, Y. Hu, Z. Song, T. Wu, F. Song, et al., 
``Non-Abelian gauge fields in circuit systems,''
Nature Electronics \textbf{5}, 635 (2022).

\bibitem{YangYang}
Y.~Yang, B.~Yang, G.~Ma, J.~Li, S.~Zhang and C.~T.~Chan,
``Non-Abelian physics in light and sound,''
Science, 383, 844-858 (2024).
DOI: 10.1126/science.adf9621
[arXiv:2305.12206 [physics.optics]].

\bibitem{NABA-PR}
Huanhuan Yang, Lingling Song, Yunshan Cao, Peng Yan,
``Circuit realization of topological physics,'' 
Physics Reports, \textbf{1093}, (2024) 1-54 
https://doi.org/10.1016/j.physrep.2024.09.007.


\bibitem{GoddardOlive}
P.~Goddard and D.~I.~Olive,
``New Developments in the Theory of Magnetic Monopoles,''
Rept. Prog. Phys. \textbf{41} (1978), 1357
doi:10.1088/0034-4885/41/9/001
 
\bibitem{Asorey81}
M.~Asorey,
``Some Remarks on the Classical Vacuum Structure of Gauge Field Theories,''
J. Math. Phys. \textbf{22} (1981), 179
[erratum: J. Math. Phys. \textbf{25} (1984), 187]
doi:10.1063/1.524732

\bibitem{Asorey82}
M.~Asorey,
``REGULARITY OF GAUGE EQUIVALENCE IN QUANTUM MECHANICS AND THE AHARONOV-BOHM EFFECT,''
Lett. Math. Phys. \textbf{6} (1982), 429-435
doi:10.1007/BF00405862

\bibitem{Kerner68}
R.~Kerner,
``Generalization of the Kaluza-Klein Theory for an Arbitrary Nonabelian Gauge Group,''
Ann. Inst. H. Poincare Phys. Theor. \textbf{9} (1968), 143-152

\bibitem{Wong70}
S. K. Wong,
``Field and particle equations for the classical Yang-Mills field and particles with isotopic spin,''
Nuovo Cimento {\bf 65A}, 689 (1970)

\bibitem{Trautman70}
A.~Trautman,
``Fiber bundles associated with space-time,''
Rept. Math. Phys. \textbf{1} (1970), 29-62
doi:10.1016/0034-4877(70)90003-0

\bibitem{Cho75}
Y.~M.~Cho,
``Higher - Dimensional Unifications of Gravitation and Gauge Theories,''
J. Math. Phys. \textbf{16} (1975), 2029
doi:10.1063/1.522434

\bibitem{Witten81}
E.~Witten,
``Search for a Realistic Kaluza-Klein Theory,''
Nucl. Phys. B \textbf{186} (1981), 412
doi:10.1016/0550-3213(81)90021-3

\bibitem{Balachandran76}
A.~P.~Balachandran, P.~Salomonson, B.~S.~Skagerstam and J.~O.~Winnberg,
``Classical Description of Particle Interacting with Nonabelian Gauge Field,''
Phys. Rev. D \textbf{15} (1977), 2308-2317
doi:10.1103/PhysRevD.15.2308

\bibitem{Balachandran77}
A.~P.~Balachandran, S.~Borchardt and A.~Stern,
``Lagrangian and Hamiltonian Descriptions of Yang-Mills Particles,''
Phys. Rev. D \textbf{17} (1978), 3247
doi:10.1103/PhysRevD.17.3247

\bibitem{Sternberg77}
S.~Sternberg,
``Minimal Coupling and the Symplectic Mechanics of a Classical Particle in the Presence of a Yang-Mills Field,''
Proc. Nat. Acad. Sci. \textbf{74} (1977), 5253-5254
doi:10.1073/pnas.74.12.5253

\bibitem{Sternberg78}
V.~Guillemin and S.~Sternberg,
``On the Equations of Motion of a Classical Particle in a {Yang-Mills} Field and the Principle of General Covariance,''
Hadronic J. \textbf{1} (1978), 1
TAUP-655-78.

\bibitem{DuvalAix79}
C. Duval, 
``On the prequantum description of spinning particles in an external gauge field,''
Proc. Aix Conference on
{\it Diff. Geom. Meths. in Math. Phys.}
Ed. Souriau. Springer LNM {\bf 836}, 49-67 (1980).
This paper is an extended version of C. Duval's unpublished
paper \cite{Duval78}.

\bibitem{Duval78}
C. Duval,
``Sur les mouvements classiques dans un champs de Yang-Mills,''
 Marseille preprint CPT-78-P-1056.

\bibitem{DH82}
C.~Duval and P.~Horvathy,
``Particles With Internal Structure: The Geometry of Classical Motions and Conservation Laws,''
Annals Phys. \textbf{142} (1982), 10
doi:10.1016/0003-4916(82)90226~-~3

\bibitem{Wipf85}
A.~W.~Wipf,
``NONRELATIVISTIC YANG-MILLS PARTICLES IN A SPHERICALLY SYMMETRIC MONOPOLE FIELD,''
J. Phys. A \textbf{18} (1985), 2379-2384
doi:10.1088/0305-4470/18/12/034

\medskip\goodbreak


\bibitem{Feher86}
L.~G.~Feher,
``Classical motion of coloured test particles along geodesics of a Kaluza-Klein spacetime,''
Acta Phys. Hung. \textbf{59} (1986), 437-444

\bibitem{HPAAix79}
P.~A.~Horvathy,
``Classical Action, the {Wu-Yang} Phase Factor and Prequantization,''
Lect. Notes Math. \textbf{836} (1980), 67-90
doi:10.1007/BFb0089727

\bibitem{WuYang69}
T. T. Wu and C. N. Yang, 
``Some solutions of the classical isotopic gauge field equations,'' in {\sl Properties of Matter under Unusual Conditions. Festschrift for the 60th birthday of E. Teller}. p. 349. Ed. H. Mark and S. Fernbach. Interscience: (1969).

\bibitem{tHooft}
G. 't Hooft,
``Magnetic monopoles in unified gauge theories,''
Nucl. Phys. \textbf{B79} (1974) 276
https://doi.org/10.1016/0550-3213(74)90486-6

\bibitem{Polyakov}
A.~M.~Polyakov,
``Particle Spectrum in Quantum Field Theory,''
JETP Lett. \textbf{20} (1974), 194-195
PRINT-74-1566 (LANDAU-INST).

\bibitem{AbbottDeser}
L.~F.~Abbott and S.~Deser,
``Charge Definition in Nonabelian Gauge Theories,''
Phys. Lett. B \textbf{116} (1982), 259-263
doi:10.1016/0370-2693(82)90338-0

\bibitem{JackiwManton}
  R.~Jackiw and N.~S.~Manton,
``Symmetries And Conservation Laws In Gauge Theories,''
  Annals Phys.\  {\bf 127} (1980) 257.
  
\bibitem{Schwarz77}
A.~S.~Schwarz,
``On Symmetric Gauge Fields,''
Commun. Math. Phys. \textbf{56} (1977), 79-86
doi:10.1007/BF01611118
   
\bibitem{ForgacsManton}
P.~Forg\'acs and N.~S.~Manton,
``Space-Time Symmetries in Gauge Theories,''
Commun. Math. Phys. \textbf{72} (1980), 15
doi:10.1007/BF01200108

\bibitem{Harnad79}
J.~P.~Harnad, L.~Vinet and S.~Shnider,
``Group Actions on Principal Bundles and Invariance Conditions for Gauge Fields,''
J. Math. Phys. \textbf{21} (1980), 2719
doi:10.1063/1.524389

\bibitem{Jackiw80}
R.~Jackiw,
``Invariance, Symmetry and Periodicity in Gauge Theories,''
Acta Phys. Austriaca Suppl. \textbf{22} (1980), 383-438
PRINT-80-0254 (SANTA-BARBARA,ITP).

\bibitem{vHolten}
J.~W.~van Holten,
``Covariant Hamiltonian dynamics,''
Phys. Rev. D \textbf{75} (2007), 025027
doi:10.1103/PhysRevD.75.025027
[arXiv:hep-th/0612216 [hep-th]].

\bibitem{HP-NGOME}
P.~A.~Horvathy and J.~P.~Ngome,
``Conserved quantities in non-abelian monopole fields,''
Phys. Rev. D \textbf{79} (2009), 127701
doi:10.1103/PhysRevD.79.127701
[arXiv:0902.0273 [hep-th]]. 

\bibitem{JackiwRebbi76}
R.~Jackiw and C.~Rebbi,
``Spin from Isospin in a Gauge Theory,''
Phys. Rev. Lett. \textbf{36} (1976), 1116
doi:10.1103/PhysRevLett.36.1116

\bibitem{Hasi}
P.~Hasenfratz and G.~'t Hooft,
``A Fermion-Boson Puzzle in a Gauge Theory,''
Phys. Rev. Lett. \textbf{36} (1976), 1119
doi:10.1103/PhysRevLett.36.1119

\bibitem{Chen:2008ag}
X.~S.~Chen, X.~F.~Lu, W.~M.~Sun, F.~Wang and T.~Goldman,
``Spin and orbital angular momentum in gauge theories: Nucleon spin structure and multipole radiation revisited,''
Phys. Rev. Lett. \textbf{100} (2008), 232002
doi:10.1103/PhysRevLett.100.232002
[arXiv:0806.3166 [hep-ph]].

\bibitem{SSD}
J.-M.~Souriau,
\textsl{Structure des syst\`emes dynamiques}, Dunod (1970). 
\textsl{Structure of Dynamical Systems. A Symplectic View of Physics}, Birkh\"auser, Boston (1997).

\bibitem{Jacobi}
C. G. J. Jacobi, 
 ``Vorlesungen \"uber Dynamik.''
Univ. K\"onigsberg 1842-43. Herausg. A. Clebsch. 
Zweite ausg. C. G. J. Jacobi's Gesammelte Werke. Supplementband. Herausg. E. Lottner. Berlin Reimer  (1884).
An english translation is available  as {\sl Jacobi's Lectures on Dynamics}, 2nd edition,
{\sl Texts and Readings in Mathematics}, https://doi.org/10.1007/978-93-86279-62-0, Hindustan Book Agency Gurgaon.

\bibitem{HP-EPL}
P.~A.~Horvathy,
``THE WU-YANG FACTOR AND THE NONABELIAN AHARONOV-BOHM EXPERIMENT,''
Eur. Phys. Lett. \textbf{2} (1986), 195
doi:10.1209/0295-5075/2/3/005%

\bibitem{HP-Kollar1}
P.~A.~Horvathy and J.~Kollar,
``The Nonabelian {Aharonov-Bohm} Effect in Geometric Quantization,''
Class. Quant. Grav. \textbf{1} (1984), L61
doi:10.1088/0264-9381/1/6/002.

\bibitem{HP-Kollar2}
P.~A.~Horvathy and J.~Kollar,
``AHARONOV-BOHM EFFECT IN SU(N) GAUGE THEORY,''
BI-TP-82-1A. (inspire code {Horvathy:1982sj}). 
In Proc. Int. Conf.  {\sl Monopoles in Quantum Field Theory}, Trieste, dec. 1981 ed. Craigie, P. Goddard, W. Nahm (eds) (1982), pp. 277-278. 

\bibitem{Sundrum}
R.~Sundrum and L.~J.~Tassie,
``Nonabelian {Aharonov-Bohm} Effects, Feynman Paths, and Topology,''
J. Math. Phys. \textbf{27} (1986), 1566-1570
doi:10.1063/1.527067

\bibitem{Arodz82}
H.~Arodz,
``Colored, Spinning Classical Particle in an External Nonabelian Gauge Field,''
Phys. Lett. B \textbf{116} (1982), 251-254
doi:10.1016/0370-2693(82)90336-7;

\bibitem{Arodz82b}
H.~Arodz,
``A Remark on the Classical Mechanics of Colored Particles,''
Phys. Lett. B \textbf{116} (1982), 255-258

\bibitem{Arodz83}:
H.~Arodz,
``LIMITATION OF THE CONCEPT OF THE CLASSICAL COLORED PARTICLE,''
Acta Phys. Polon. B \textbf{14} (1983), 13-21
%

\bibitem{Feher:1984ik} 
L.~G.~Feh\'er,
``Bounded Orbits for Classical Motion of Colored Test Particles in the Prasad-Sommerfield Monopole Field,''
Acta Phys. Polon. B \textbf{15} (1984), 919
Print-84-0247 (BOLYAI).

\bibitem{Feher:1984xc}
L.~G.~Feh\'er,
``Quantum Mechanical Treatment of an Isospinor Scalar in {Yang-Mills} Higgs Monopole Background,''
Acta Phys. Polon. B \textbf{16} (1985), 217
PRINT-84-0552 (BOLYAI-INST).

\bibitem{DeaverFairbank}
B. S. Deaver, Jr. , and W. M. Fairbank, 
``Experimental evidence for quantized flux in superconducting cylinders,'' 
Phys. Rev. Letters 7, 43 (1961)

\bibitem{ByersYang}
N.~Byers and C.~N.~Yang,
``Theoretical Considerations Concerning Quantized Magnetic Flux in Superconducting Cylinders,''
Phys. Rev. Lett. \textbf{7} (1961), 46-49
doi:10.1103/PhysRevLett.7.46

\bibitem{Onsager}
L. Onsager, 
``Magnetic Flux Through a Superconducting Ring,''
Phys. Rev. Lett. 7, 50 (1961)
DOI:https://doi.org/10.1103/PhysRevLett.7.50

\bibitem{JackiwWeinberg}
R.~Jackiw and E.~J.~Weinberg,
``SELFDUAL CHERN-SIMONS VORTICES,''
Phys. Rev. Lett. \textbf{64} (1990), 2234
doi:10.1103/PhysRevLett.64.2234

\bibitem{JackiwPiCS}
R.~Jackiw and S.~Y.~Pi,
``Classical and quantal nonrelativistic Chern-Simons theory,''
Phys. Rev. D \textbf{42} (1990), 3500
[erratum: Phys. Rev. D \textbf{48} (1993), 3929]
doi:10.1103/PhysRevD.42.3500

\bibitem{DJPT}
G.~V.~Dunne, R.~Jackiw, S.~Y.~Pi and C.~A.~Trugenberger,
``Selfdual Chern-Simons solitons and two-dimensional nonlinear equations,''
Phys. Rev. D \textbf{43} (1991), 1332
[erratum: Phys. Rev. D \textbf{45} (1992), 3012]
doi:10.1103/PhysRevD.43.1332

\bibitem{D-LNP}
G.~V.~Dunne,
``Selfdual Chern-Simons theories,''
Lect. Notes Phys. M \textbf{36} (1995), 1-217
1995,
doi:10.1007/978-3-540-44777-1

\bibitem{FrenkelHrasko}
A.~Frenkel and P.~Hrasko,
``Invariance Properties of the Dirac Monopole,''
Annals Phys. \textbf{105} (1977), 288
doi:10.1016/0003-4916(77)90242-1

\bibitem{HraskoBalog}
P.~Hrasko and J.~Balog,
``Rotation Symmetry in the Hamiltonian Dynamics,''
Nuovo Cim. B \textbf{45} (1978), 239-254
doi:10.1007/BF02894683

\bibitem{GNO}
P.~Goddard, J.~Nuyts and D.~I.~Olive,
``Gauge Theories and Magnetic Charge,''
Nucl. Phys. B \textbf{125} (1977), 1-28
doi:10.1016/0550-3213(77)90221-8

\bibitem{HRCMP1}
P.~A.~Horvathy and J.~H.~Rawnsley,
``Topological Charges in Monopole Theories,''
Commun. Math. Phys. \textbf{96} (1984), 497
doi:10.1007/BF01212532

\bibitem{MITNews}
David L. Chandler, 
``Exotic physics phenomenon is observed for first time.
Observation of the predicted non-Abelian Aharonov-Bohm Effect may offer step toward fault-tolerant quantum computers,''
https://news.mit.edu/2019/aharonov-bohm-effect-physics-observed-0905 Sept. 5 (2019).


\end{thebibliography}
\end{document}